\documentclass[twocolumn]{aastex62}
\usepackage{graphicx}
\usepackage{natbib}
 \usepackage{sidecap}
\usepackage{rotating}
\received{January 1, 2018}
\revised{January 7, 2018}
\accepted{\today}

\submitjournal{ApJ}

\shorttitle{as~209}
\shortauthors{Favre et al.}
\begin{document}

%-----------------------------------------------------------------------
% TITLE & AUTHORS
%-----------------------------------------------------------------------
%
\title{Gas density perturbations induced by forming planet(s) in the AS~209 protoplanetary disk as seen with ALMA}

\correspondingauthor{C\'ecile Favre}
\email{cfavre@arcetri.astro.it}

\author[0000-0002-5789-6931]{C\'ecile Favre}
\affil{INAF-Osservatorio Astrofisico di Arcetri, Largo E. Fermi 5, I-50125, Florence, Italy}

\author{Davide Fedele}
\affiliation{INAF-Osservatorio Astrofisico di Arcetri, Largo E. Fermi 5, I-50125, Florence, Italy}

\author{Luke Maud}
\affiliation{Leiden Observatory, Leiden University, PO Box 9513, 2300 RA Leiden, The Netherlands}

\author{Richard Booth}
\affiliation{Institute of Astronomy, University of Cambridge, Madingley Road, Cambridge CB3 0HA, UK}

\author{Marco Tazzari}
\affiliation{Institute of Astronomy, University of Cambridge, Madingley Road, Cambridge CB3 0HA, UK}

\author{Anna Miotello}
\affiliation{Leiden Observatory, Leiden University, PO Box 9513, 2300 RA Leiden, The Netherlands}

\author{Leonardo Testi}
\affiliation{European Southern Observatory, Karl-Schwarzschild-Str. 2, 85748 Garching, Germany}
\affiliation{Excellence Cluster Universe, Boltzmannstr. 2, 85748 Garching, Germany}
\affiliation{INAF-Osservatorio Astrofisico di Arcetri, Largo E. Fermi 5, I-50125, Florence, Italy}

\author{Dmitry Semenov}
\affiliation{Max Planck Institute for Astronomy, Königstuhl 17, 69117 Heidelberg, Germany}
\affiliation{Department of Chemistry, Ludwig Maximilian University, Butenandtstr. 5-13, D-81377 Munich, Germany}

\author{Simon Bruderer}
\affiliation{Max Planck Institut f\"ur Extraterrestrische Physik, Giessenbachstrasse 1, 85748 Garching, Germany}

%===================================================================================================
%===================================================================================================
%
%-----------------------------------------------------------------------
% ABSTRACT
%-----------------------------------------------------------------------
%

\begin{abstract}

The formation of planets occurs within protoplanetary disks surrounding young stars, resulting in perturbation of the gas and dust surface densities. Here, we report the first  evidence of spatially resolved gas surface density ($\Sigma_{g}$) perturbation towards the AS~209 protoplanetary disk from the optically thin C$^{18}$O ($J=2-1$) emission. The observations were carried out at 1.3~mm with ALMA at a spatial resolution of about 0.3$\arcsec$ $\times$ 0.2$\arcsec$ (corresponding to $\sim$ 38 $\times$ 25 au). The C$^{18}$O emission shows a compact ($\le$60~au), centrally peaked emission and an outer ring peaking at 140~au, consistent with that observed in the continuum emission and, its azimuthally averaged radial intensity profile presents a deficit that is spatially coincident with the previously reported dust map. This deficit can only be reproduced with our physico-thermochemical disk model by lowering $\Sigma_{gas}$ by nearly an order of magnitude in the dust gaps. Another salient result is that contrary to C$^{18}$O, the DCO$^{+}$ ($J=3-2$) emission peaks between the two dust gaps. We infer that the best scenario to explain our observations (C$^{18}$O deficit and DCO$^{+}$ enhancement) is a gas perturbation due to forming-planet(s), that is commensurate with previous continuum observations of the source along with hydrodynamical simulations. Our findings confirm that the previously observed dust gaps are very likely due to perturbation of the gas surface density that is induced by a planet of at least 0.2~M$\rm_{Jupiter}$ in formation. Finally, our observations also show the potential of using CO isotopologues to probe the presence of saturn mass planet(s).
\end{abstract}

%-----------------------------------------------------------------------
% KEYWORDS
%-----------------------------------------------------------------------
%
%
      
%===============================================================
%===============================================================

%-----------------------------------------------------------------------
%----------INTRODUCTION -------------
%-----------------------------------------------------------------------
\section{Introduction} \label{sec:intro}

%-----------------------------------------------------------------------------------------------
% --- TABLE 1 ---
%-----------------------------------------------------------------------------------------------
\begin{table*}
\begin{center}
\caption{\label{tab1}
Spectroscopic and Observational line parameters.}
\begin{tabular}{llccccccc} 
\hline \hline
 Molecule & Transition & Frequency &	E$\rm_{up}$ 	& S$\mu$$^{2}$ & \multicolumn{2}{c}{Synthesized beam} & rms & Integrated Flux\tablenotemark{a}	\\
         &       & (MHz) & (K) & (D$^{2}$)  & ($\arcsec$ $\times$ $\arcsec$) & PA ($\degr$) & (mJy~beam$^{-1}$) & (K~km~s$^{-1}$)\\
\hline
CO &(2--1) & 230538.0 &16.6 & 0.02 & 0.25 $\times$ 0.21 & $-$75.85 & 3.5 & 27.1 $\pm 3.9$\\
$^{13}$CO  &(2--1) \tablenotemark{b}&220398.7& 15.9& 0.05 &  0.25 $\times$ 0.21 &$-$73.43 & 3.5& 8.5 $\pm$ 2.6\\
C$^{18}$O &(2--1) &219560.4 &15.8 & 0.02 & 0.26 $\times$ 0.22 &$-$72.41 & 2.7& 3.4 $\pm$ 2.1 \\
DCO$^{+}$ &(3--2)  \tablenotemark{b} &216112.6&20.7& 142 &0.26 $\times$ 0.21 &$-$73.87& 2.6& 2.6 $\pm$ 2.0\\
\hline
\hline
\end{tabular}
\end{center}
\tablenotetext{a}{Measured disk-averaged over 200~au line integrated intensity.}
\tablenotetext{b}{Hyperfine splitting.}
\tablecomments{We used the spectroscopic data parameters from \citet{Caselli:2005} for DCO$^{+}$, from \citet{Winnewisser:1985}, \citet{Goorvitch:1994}, \citet{Winnewisser:1997}, \citet{Cazzoli:2004} and \citet{Klapper:2001} and for CO and its isotopologues.}
\end{table*}
%++++++++++++++++++++++++++++++++++++++++++++++++++++++++++++++++++

Formation of planets occurs within the gaseous and dusty interior of protoplanetary disks during the early phase of star formation. Although it remains difficult to directly detect planets in formation, the effect of the planet--disk interaction makes the indirect detection possible via the means of observations of the dust and molecular content of protoplanetary disks. Indeed, while forming, a planet will open gap(s) within the disk leading to a localised deficit in dust and gas \citep[e.g.,][]{Papaloizou:1984,Paardekooper:2004,Durmann:2015,Dong:2015,Dong:2017,Rosotti:2016,Bae:2017,Bae:2018}. This will result into dust and molecular gaps and rings. In that light, thanks to the high resolution and sensitivity of the Atacama Large Millimeter/submillimeter Array (ALMA) and the Spectro-Polarimetric High-contrast Exoplanet (SPHERE) facility, such structures have been observed in continuum emission \citep[e.g.,][]{ALMA-Partnership:2015,Isella:2016,Andrews:2016,Zhang:2016,Loomis:2017,van-Boekel:2017,Fedele:2017,Fedele:2018,Muro-Arena:2018}. 

Using ALMA observations, \citet{Teague:2017} has reported oscillatory features in CS emission towards TW~Hya and \citet{Harsono:2018} have reported a deficit of CO isotopologue emission within the 15~au of the disk surrounding TMC1A. Those deficits might be due to grains size and growth propoperties but also to planet(s) in formation.

Recently, using ALMA observation of CO, \citet{Teague:2018} and \citet {Pinte:2018}, claimed the first kinematic evidences of embedded forming-planet(s) in the protoplanetary disk surrounding  the Herbig Ae star HD163296. Interestingly enough, hydrodynamical models predict that not only the kinematics but also the bulk of the emission of CO isotopologues can be used as an indirect probe of the planet-disk interaction \citep[e.g.,][]{Ober:2015,Facchini:2018}; which induces a gap in the gas at the planet location.

In this paper, we investigate gas perturbations caused by a forming-planet in the disk surrounding the T Tauri star AS~209 \citep{Andrews:2009,Huang:2016,Huang:2017,Fedele:2018,Teague:2018b}, which is located at 126~pc from the Sun \citep{Gaia-Collaboration:2016}. 
For this purpose, we follow up the observations by \citet[][hereafter, Paper I]{Fedele:2018}, in which the authors inferred that the observed rings and gaps of the 1.3~mm continuum emission are likely due to the formation of planets; this hypothesis is commensurate with hydrodynamical simulations (for further details, see Paper I).
In this study, we only focus on the emission lines of CO and its isotopologues $^{13}$CO and C$^{18}$O along with that of DCO$^{+}$. In Section~\ref{sec:observations}, we briefly present the observations and the methodology used for combining ALMA observations performed during different Cycles. Results and modelling are presented in Sections~\ref{results} and ~\ref{modeling}, respectively; and discussed in Section~\ref{discussion}.

%--------------------------------------------------------------------
%--------- Observations & data reduction ---------
%--------------------------------------------------------------------
\section{Observations and data reduction}
\label{sec:observations}
%--------------------------------------------------------------------
%
The observations of the CO (2$-$1), $^{13}$CO (2$-$1), C$^{18}$O (2$-$1) and DCO$^{+}$ (3$-$2) lines (see spectroscopic parameters in Table~\ref{tab1}) were performed with ALMA towards AS~209 with 38 antennas on 2016 September 22 and 41 antennas on 2016 September 26 {\it (project ID ALMA$\#$2015.1.00486.S, PI: D. Fedele)} towards the following phase-tracking center ($\rm\alpha_{J2000}$=16$\rm^{h}49^{m}15\fs296$, $\rm\delta_{J2000}$= $-$14$\degr22\arcmin09\farcs$02). The observations cover the frequency range 211--275~GHz in band 6. For further details, see Paper~I. 

To optimize the $(uv)$-coverage together with the sensitivity, we have combined our data with the ones observed by \citet{Huang:2016} with ALMA during its Cycle 2 {\it (project ID ALMA$\#$2013.1.00226.S, PI: D. K.~\"Oberg)}. The latter were taken in relatively unstable conditions for the phase, and were shifted with respect to our observations. Therefore, the datasets were first individually self-calibrated before being merged, allowing us to properly shift the Cycle~2 observations (via the use of a "false" model point) at the same coordinate position as our data.

We use the Common Astronomy Software Applications \citep[CASA,][]{McMullin:2007} software version 4.7.2 for data reduction, selfcalibration, continuum subtraction, and the version 5.1.1 for  deconvolution and imaging. To improve the signal-to-noise ratio, we use a natural cleaning. 
The resulting synthesized beams for each molecule are given in Table~\ref{tab1} and the spectral resolution is 0.2~km~s$^{-1}$.

%------------------------------------------------------------------
% --- FIGURE 1 ---
%-----------------------------------------------------------------
\begin{figure*}
\centering
% trim={<left> <lower> <right> <upper>}
\includegraphics[trim={0 4cm 0 4cm},angle=0,width=4.4cm]{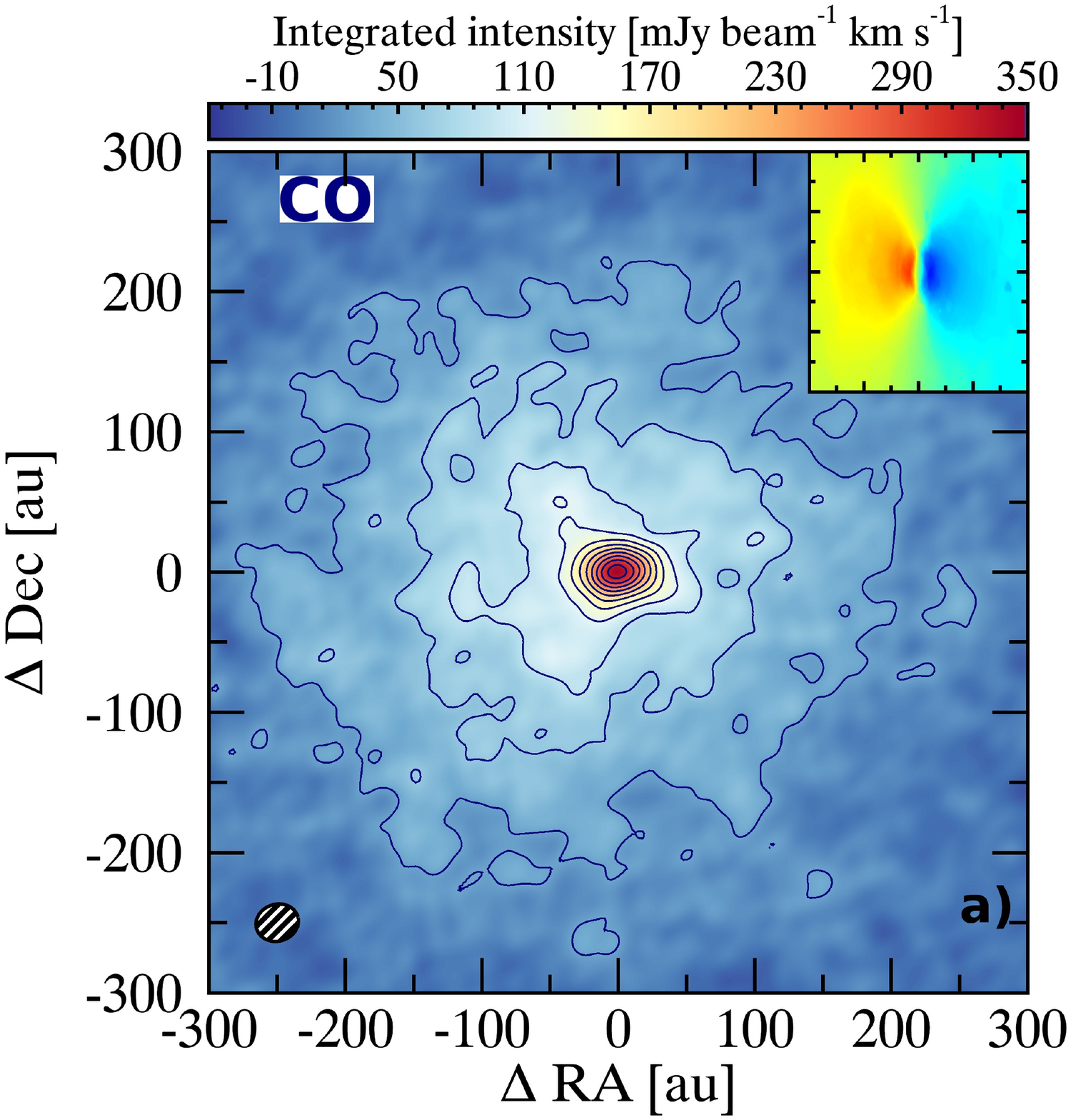}
\includegraphics[trim={0 4cm 0 4cm},angle=0,width=4.4cm]{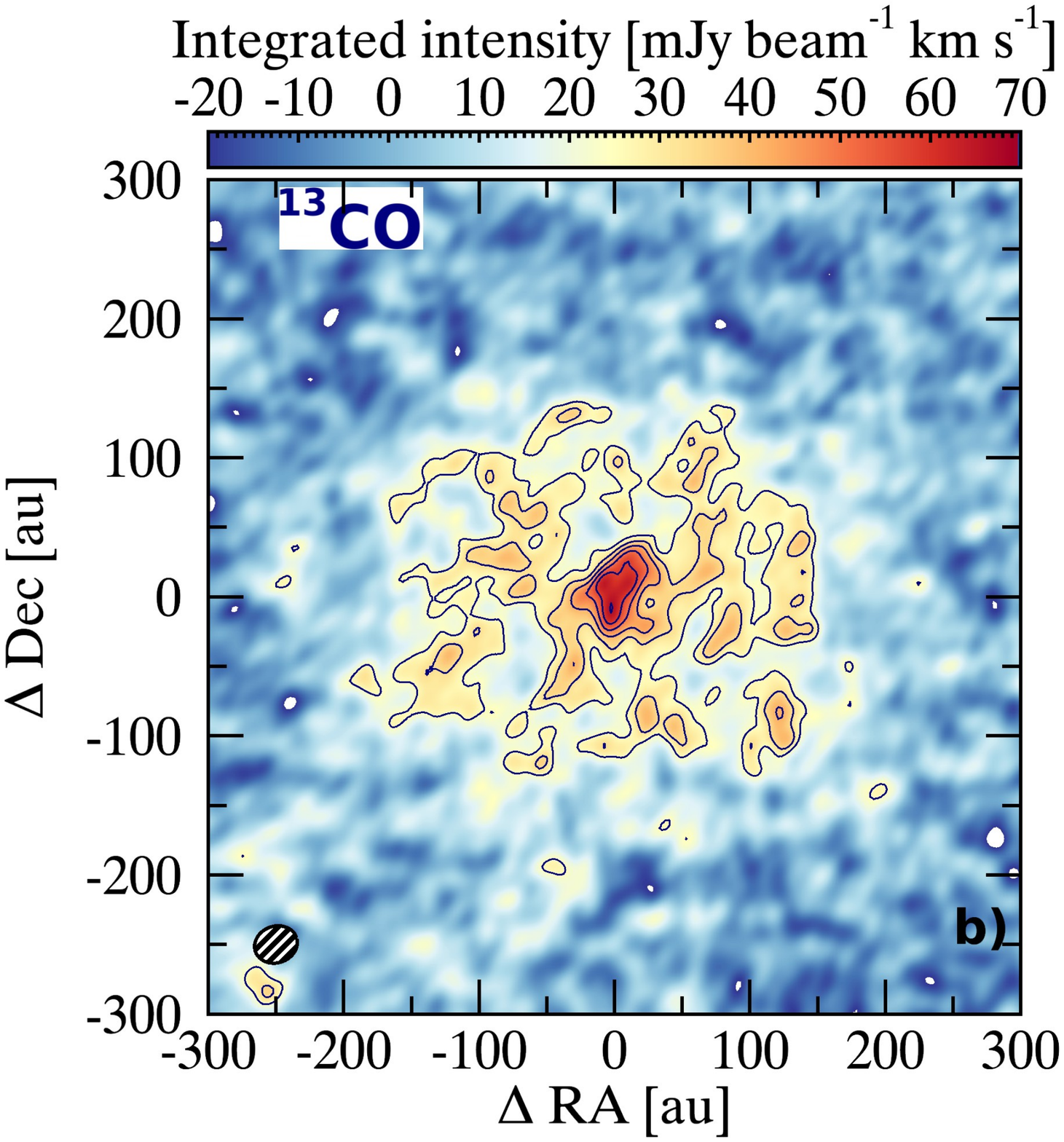}
\includegraphics[trim={0 4cm 0 4cm},angle=0,width=4.4cm]{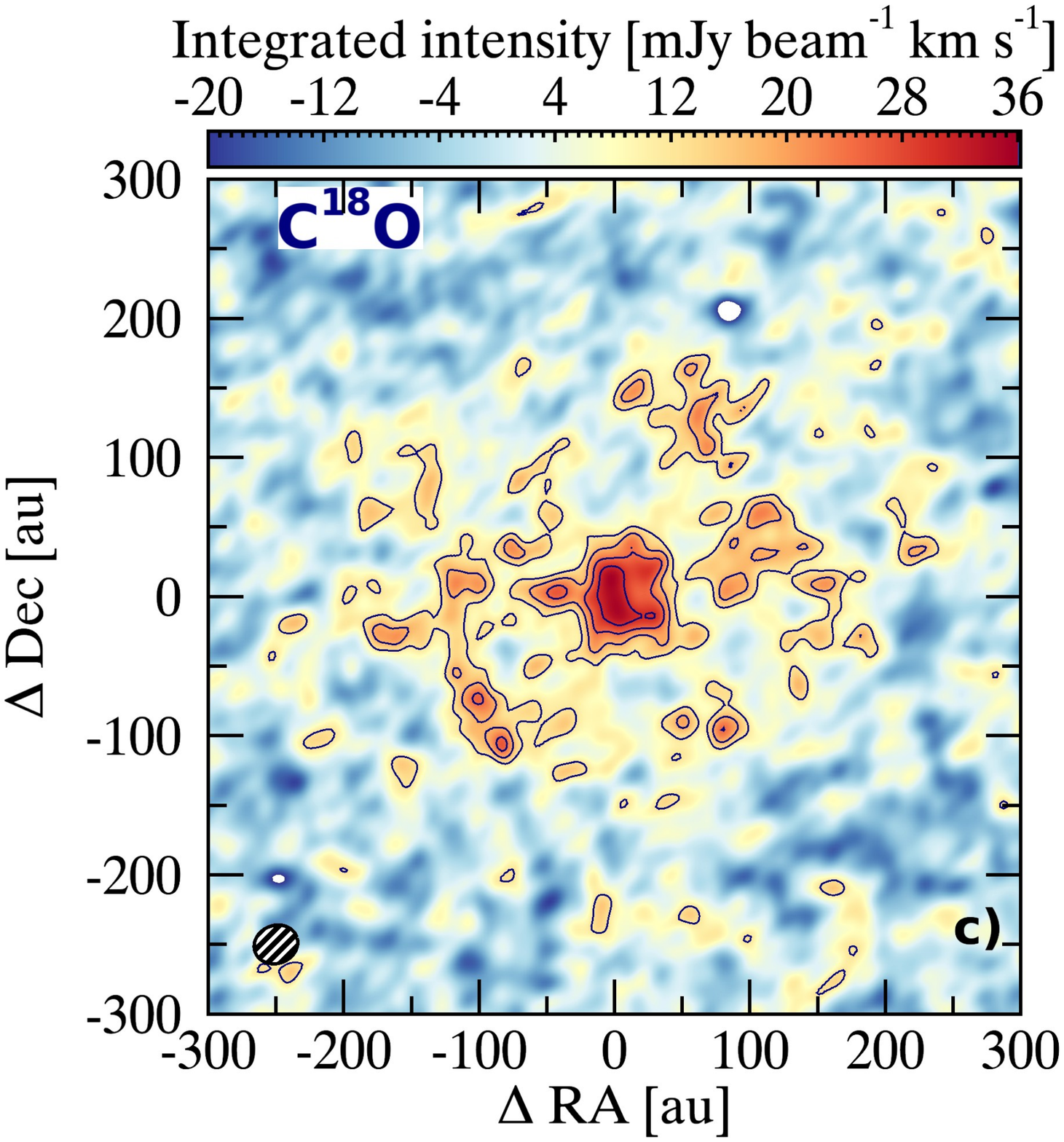}
\includegraphics[trim={0 4cm 0 4cm},angle=0,width=4.4cm]{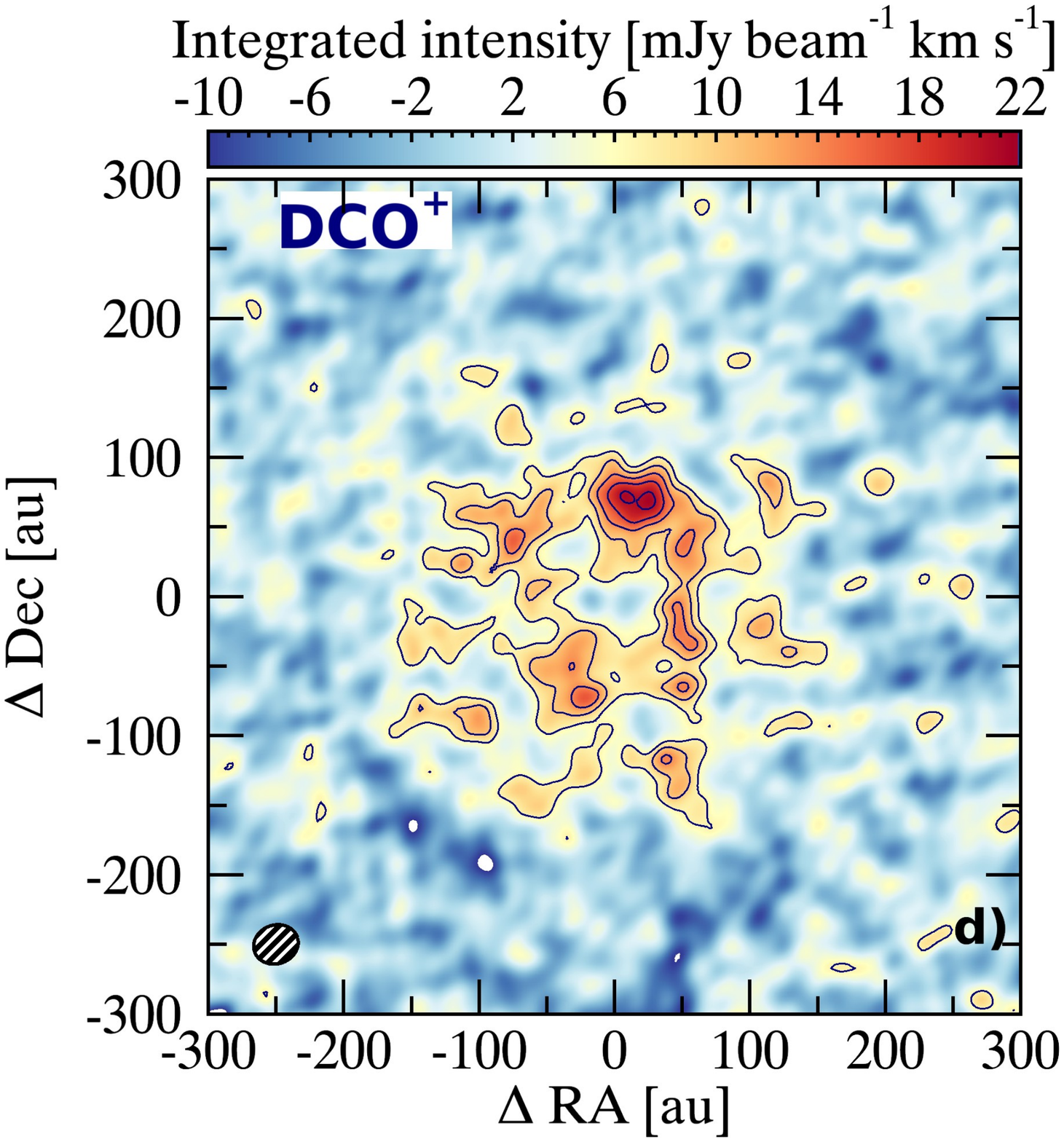}
\includegraphics[trim={0 4cm 0 4cm},angle=0,width=4.4cm]{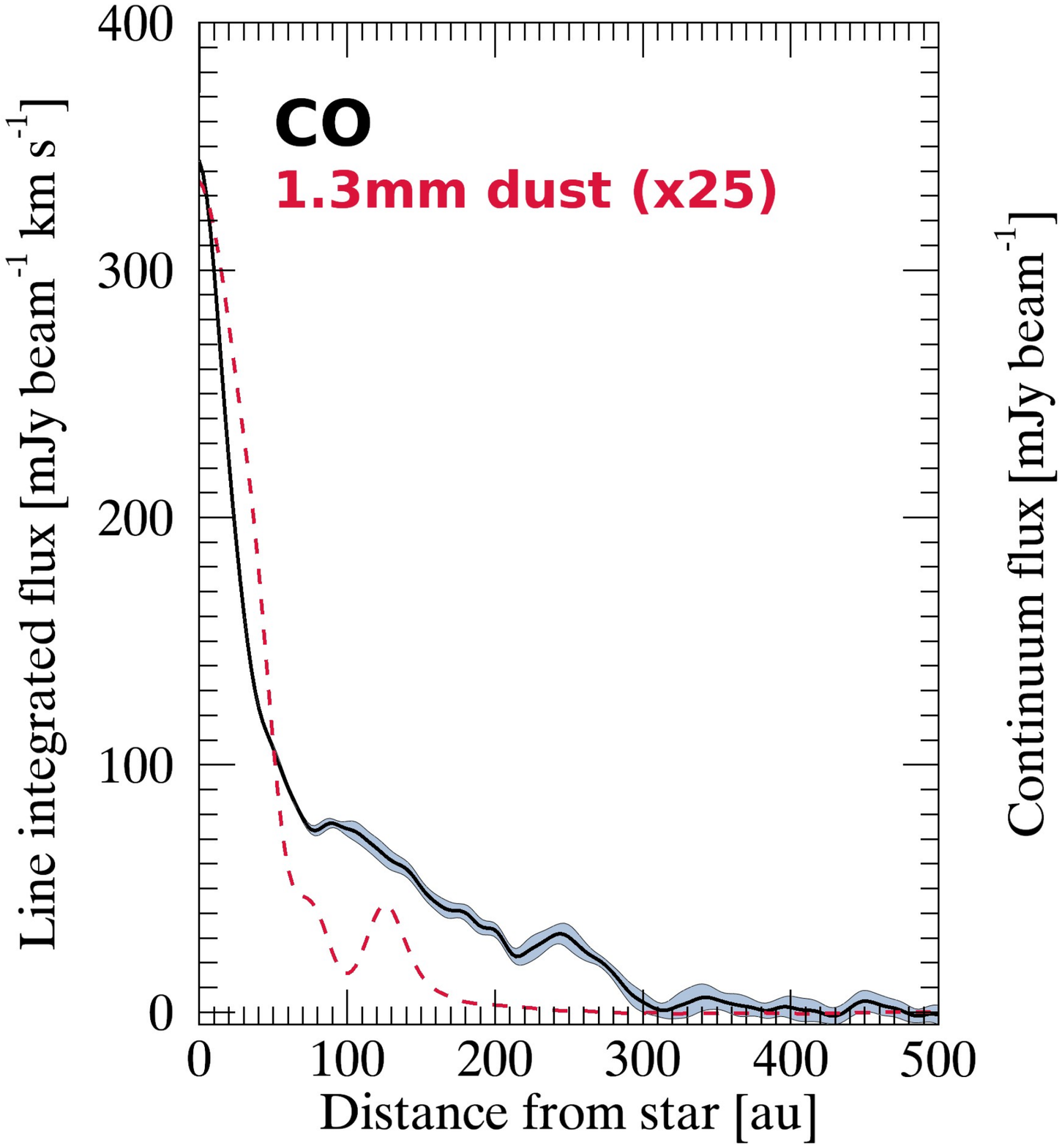}
\includegraphics[trim={0 4cm 0 4cm},angle=0,width=4.4cm]{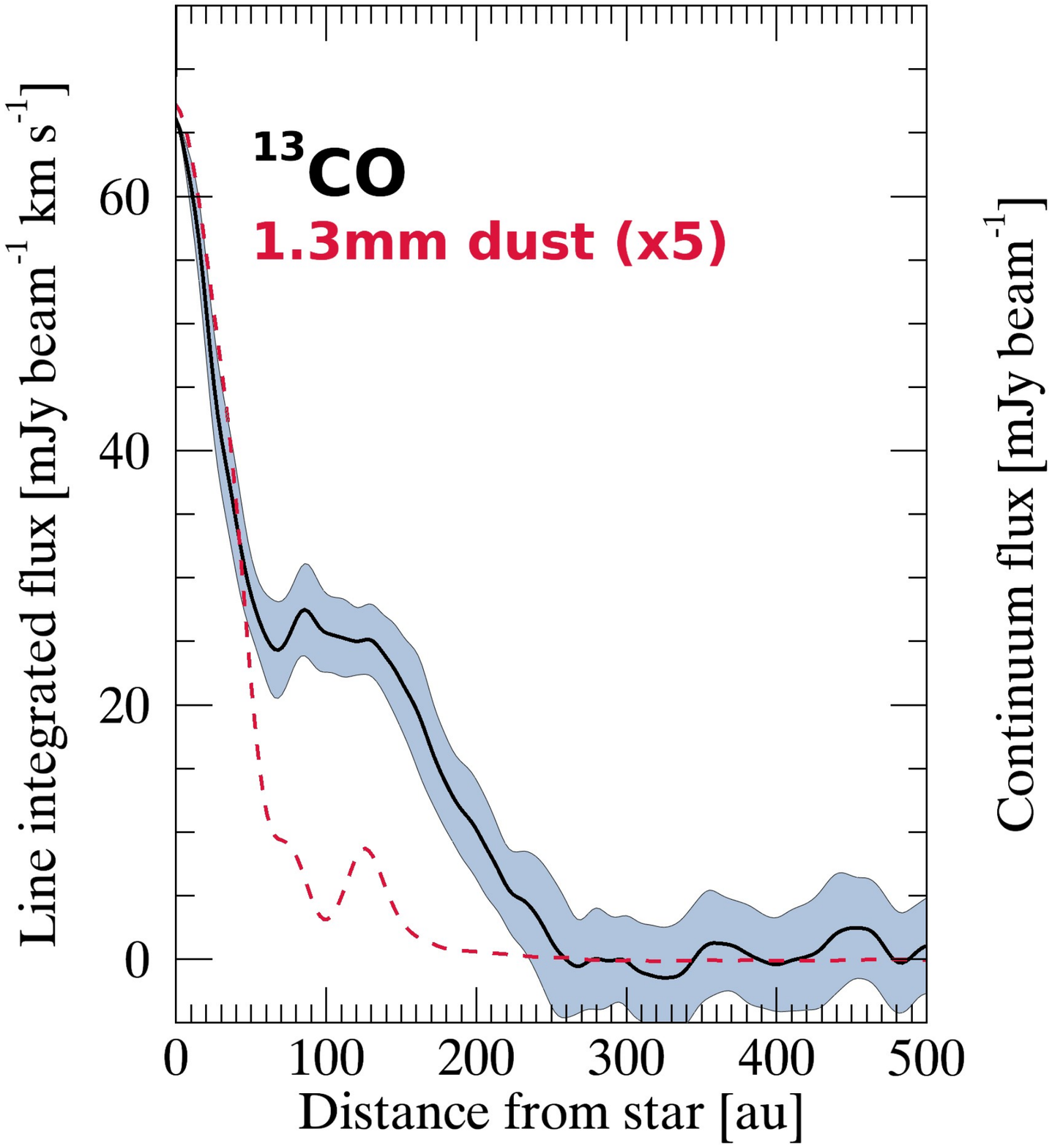}
\includegraphics[trim={0 4cm 0 4cm},angle=0,width=4.4cm]{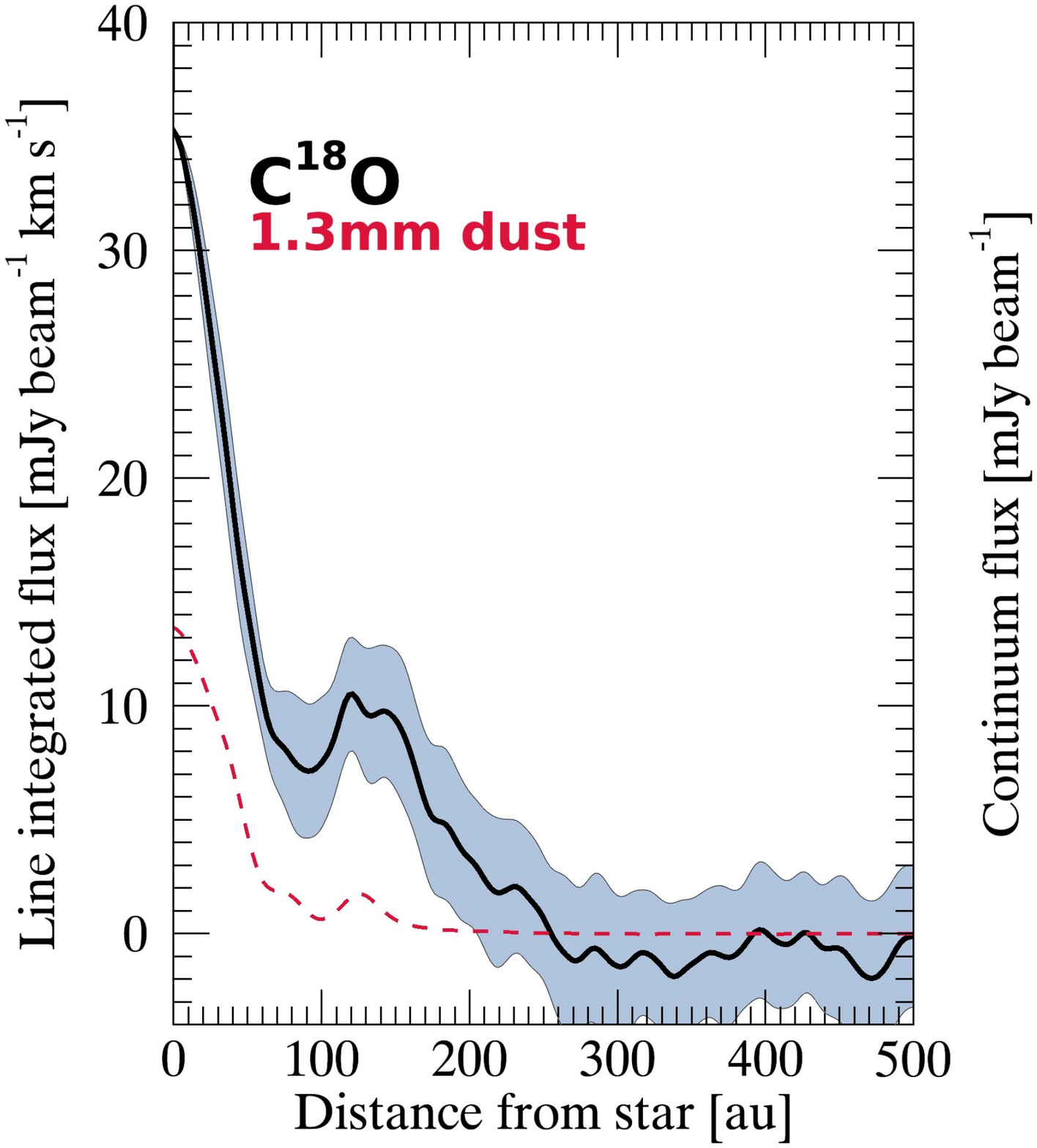}
\includegraphics[trim={0 4cm 0 4cm},angle=0,width=4.4cm]{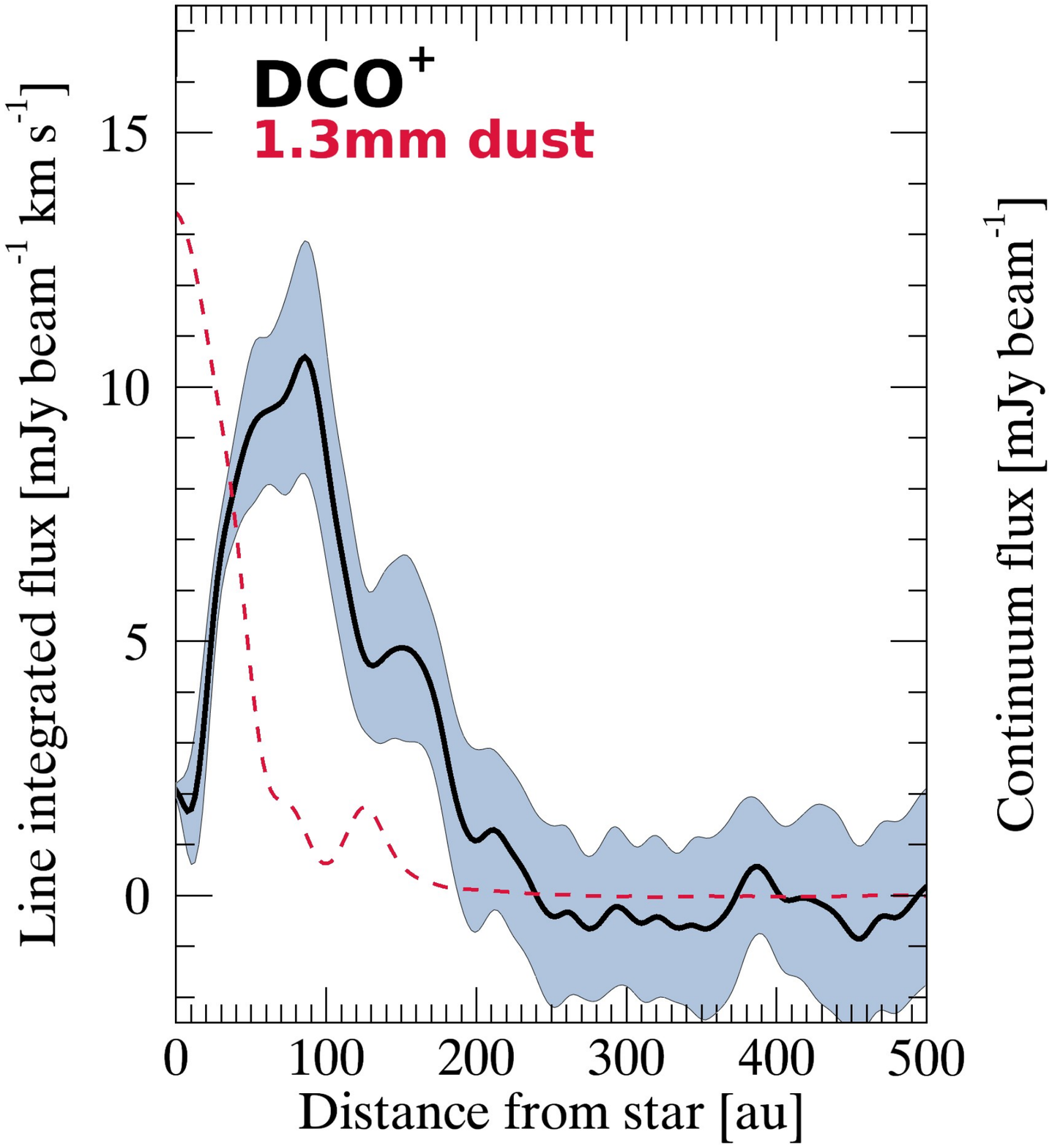}
\caption{AS~209 emission maps and radial intensity profiles.
{\it Top row:} (a) CO integrated emission map from $v_{LSR}$=$-$3.4 up to 12.0~km~s$^{-1}$. The inset shows the CO intensity-weighted mean velocity map. (b) and (c) $^{13}$CO and C$^{18}$O integrated emission maps from $v_{LSR}$=$-$1.6 up to 11.6~km~s$^{-1}$. (d) DCO$^{+}$ integrated emission map from $v_{LSR}$=1.8 up to 7.8~km~s$^{-1}$. The first contour is at 2$\sigma$ and the level step is at 1$\sigma$ for DCO$^{+}$ and C$^{18}$O (where 1$\sigma$= 3.5 and 6.2~mJy~beam$^{-1}$~km~s$^{-1}$, respectively) while for $^{13}$CO and CO, the first contour is at 3$\sigma$ and the level step is at 1$\sigma$ and 3$\sigma$, respectively (where 1$\sigma$=8.4 and 10~mJy~beam$^{-1}$~km~s$^{-1}$ for $^{13}$CO and CO, respectively). The synthesized beam is shown in the bottom-left corner of each panels.
{\it Bottom row, from left to right:} Continuum-subtracted CO, $^{13}$CO, C$^{18}$O and DCO$^{+}$ radial intensity line emission profiles (black) overlaid with that of the continuum (dash red). On each plots, the black line shows the mean profile while the shadowed regions shows the standard deviation along the azimuth angle. After deprojecting for the AS~209 disk inclination ($i$= 35$\degr$, see Paper~I), all the profiles were azimuthally averaged ($PA$= 86$\degr$), except that of CO which was averaged on a wedge $\pm$20$\degr$ along the major emission axis.
 }
\label{fg1}
\end{figure*}

%-----------------------------------------------------------------
%-----  Results and analysis-----------
%-----------------------------------------------------------------
\section{Results and analysis}
\label{results}
%-----------------------------------------------------------------

\subsection{Emission maps and velocity structure}
The CO, $^{13}$CO, C$^{18}$O and DCO$^{+}$ integrated emission maps over the line profile and the CO velocity map (that is consistent with Keplerian rotation) are displayed in Figure~\ref{fg1}. The CO emission extends beyond that of 1.3~mm dust continuum  (i.e. r$\ge$200 au), as previously observed by \citet{Huang:2016}. The asymmetry seen in the CO emission map is due to the fact that part of the emission is absorbed by the cloud \citep[see][]{Oberg:2011d}.

One notable feature shown in Figure~\ref{fg1} is that the spatial distribution of both the CO and $^{13}$CO emission is centrally peaked while that of C$^{18}$O and DCO$^{+}$ emission present rings. Indeed, C$^{18}$O  displays both a centrally bright emission inside the inner $\sim$50~au of the disk and an outer ring located at $R > 120\,$au just after the outer continuum gap reported in Paper I. Regarding the DCO$^{+}$ emission, two rings are observed: the first one located between the two dust continuum gaps (i.e., 66 au $\lessapprox$ $R$ $\lessapprox$ 95 au) and the second one lying after the outer continuum gap at $R > 120\,$au.

Figures~\ref{fg8} to~\ref{fg11}, in Appendix, display the CO, $^{13}$CO, C$^{18}$O and DCO$^{+}$ channel emission maps, respectively. The fainter lines, associated with C$^{18}$O and DCO$^{+}$, are detected with at least a peak to noise ratio $\ge$6$\sigma$ in several channels (i.e. more than 10 channels). The pattern of the CO, $^{13}$CO, C$^{18}$O and DCO$^{+}$ emission is consistent with gas in Keplerian rotation motion. 

%------------------------------------------------------------------
% --- FIGURE 2---
%-----------------------------------------------------------------
\begin{figure}
\centering
% trim={<left> <lower> <right> <upper>}
\includegraphics[trim={0 5cm 1cm 0},clip,angle=270,width=10cm]{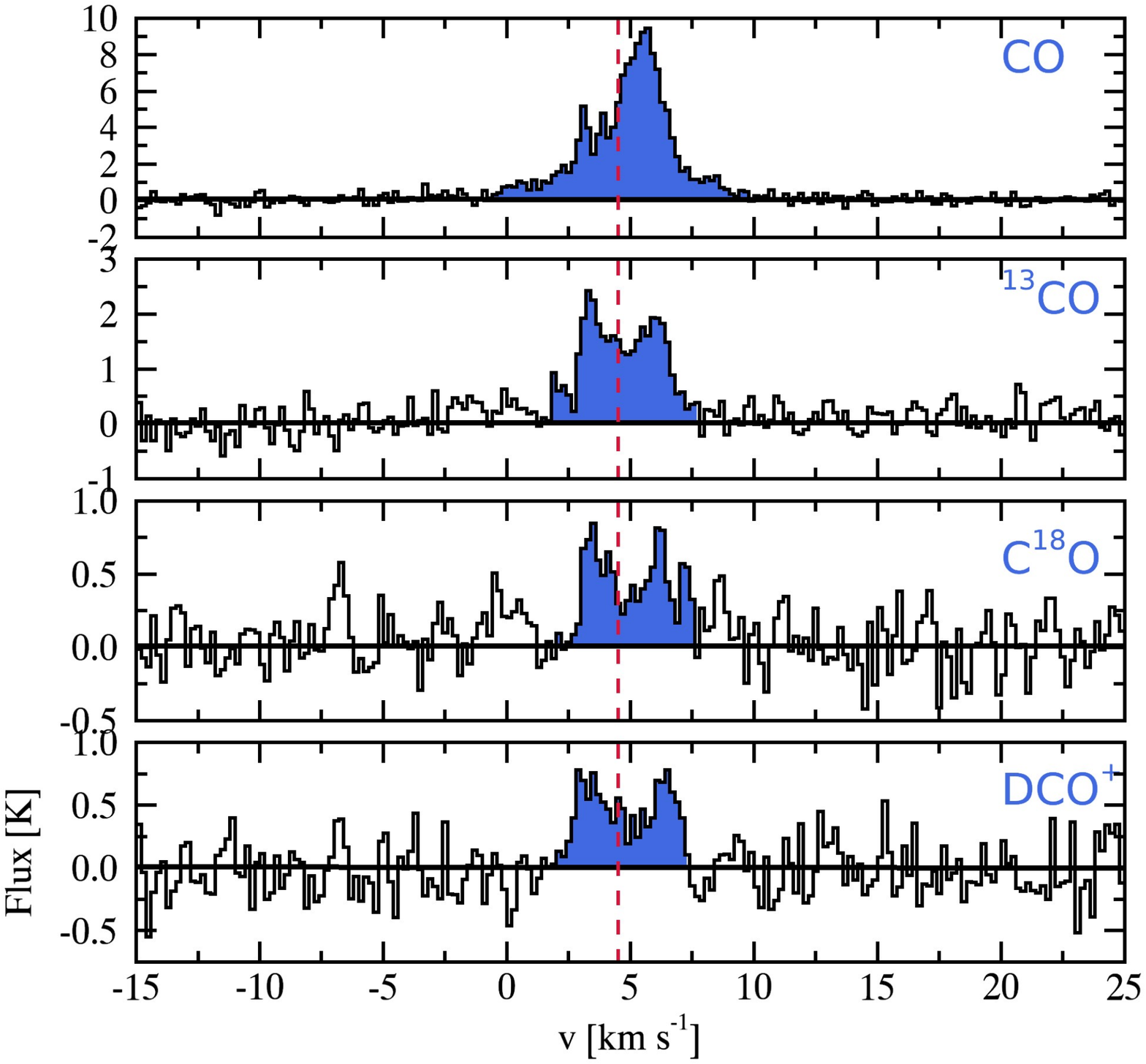}
\caption{AS~209 disk-averaged spectrum extracted within 200~au for CO, $^{13}$CO, C$^{18}$O and DCO$^{+}$. The vertical red dashed line indicated the LSR systemic velocity, $v_{LSR}\simeq$4.5~km~s$^{-1}$ of the source.}
\label{fg2}
\end{figure}

\subsection{Spectra}
Figure~\ref{fg2} displays the spatially integrated CO, $^{13}$CO, C$^{18}$O and DCO$^{+}$ spectra  extracted from a 200~au (i.e. $\sim$1.6$\arcsec$) box centered on AS~209. The line profiles are all consistent with the ${LSR}$ velocity of the source. In addition, all molecular line shapes are consistent with each other, except that of CO that displays a brighter red-shifted emission. The latter is the result of absorption by the ambient cloud of part of the blue-shifted CO emission (see above as well as Figs.~\ref{fg1} and \ref{fg3}).
\newpage

\subsection{Line opacity}
The estimate of the molecular surface density is strongly affected by the opacity of the targeted transition \citep[e.g.][]{Pietu:2007}. Assuming isotopic ratios for the local ISM of $^{16}$O/$^{18}$O = 557 and $^{12}$C/$^{13}$C=70 \citep[][]{Wilson:1999}, we estimate that the C$^{18}$O emission is optically thin throughout the 200~au disk radii (with $\tau$(C$^{18}$O) $\le$ 0.8) while that of $^{13}$CO and CO is optically thick with $\tau$ $\ge$ 2.5.

\begin{figure}
\centering
% trim={<left> <lower> <right> <upper>}
\includegraphics[trim={0 4cm 0 4cm},angle=0,width=4.2cm]{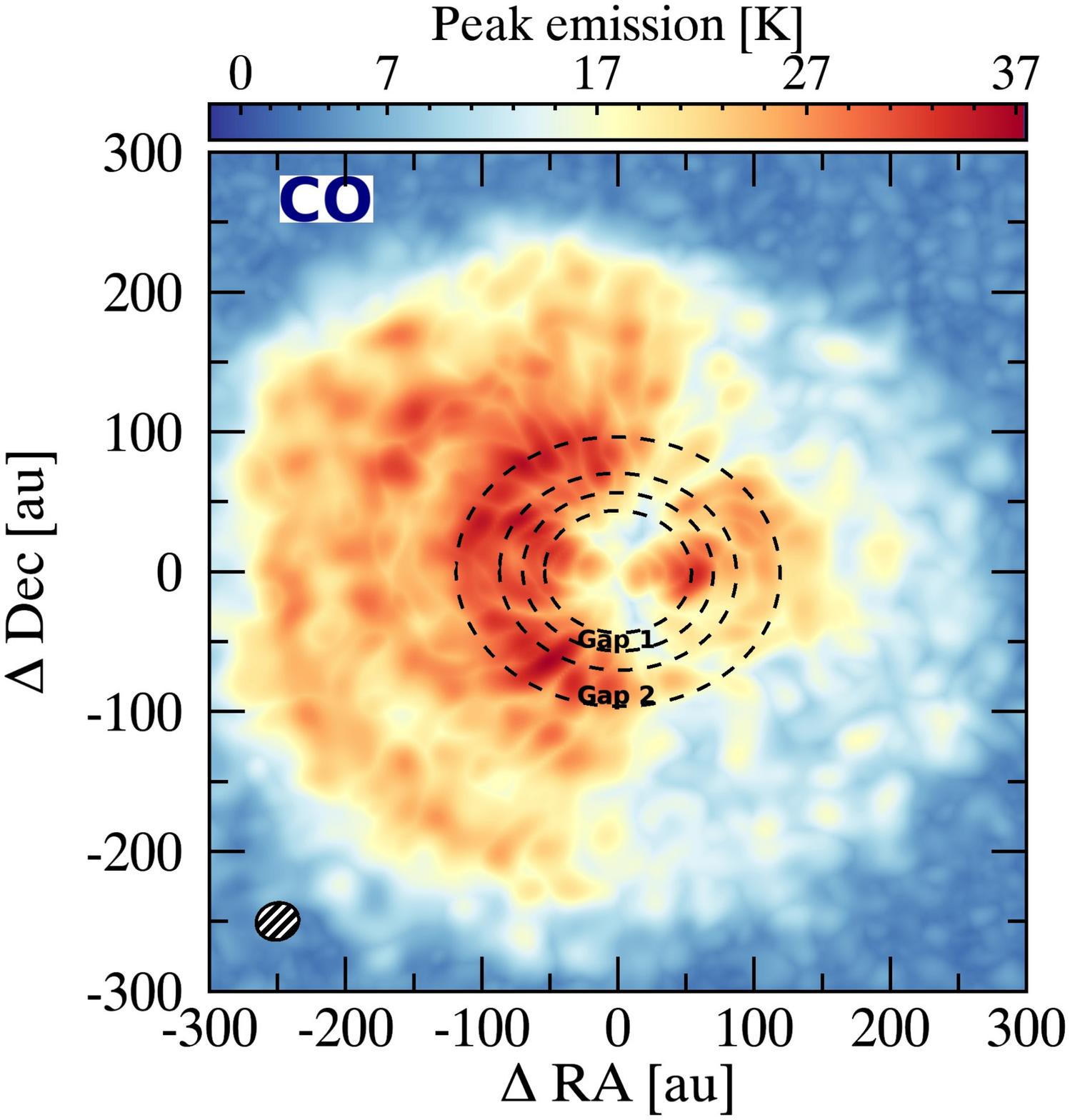}
\includegraphics[trim={0 4cm 0 4cm},angle=0,width=4.2cm]{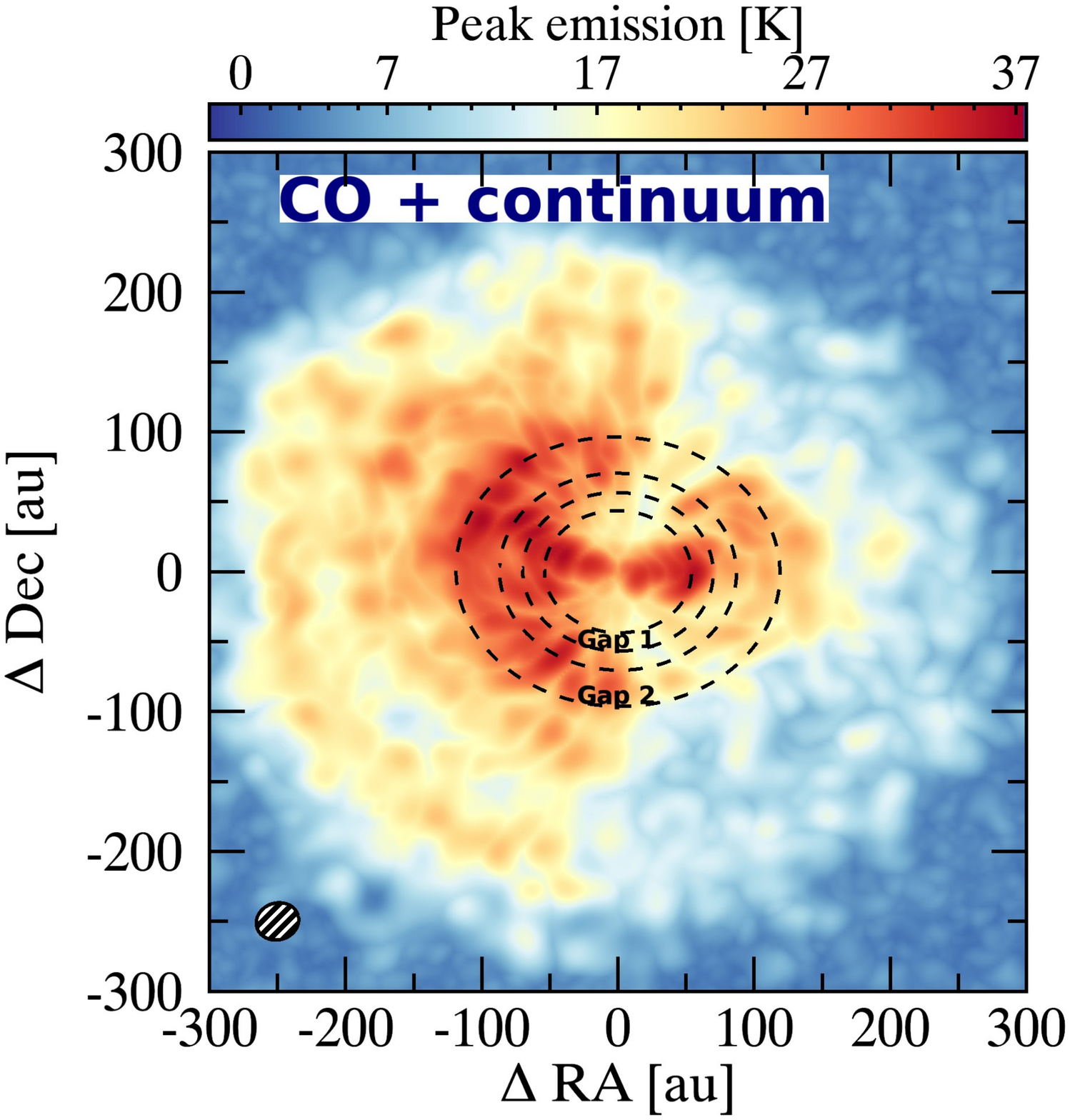}

\includegraphics[trim={0 4cm 0 4cm},angle=0,width=4.2cm]{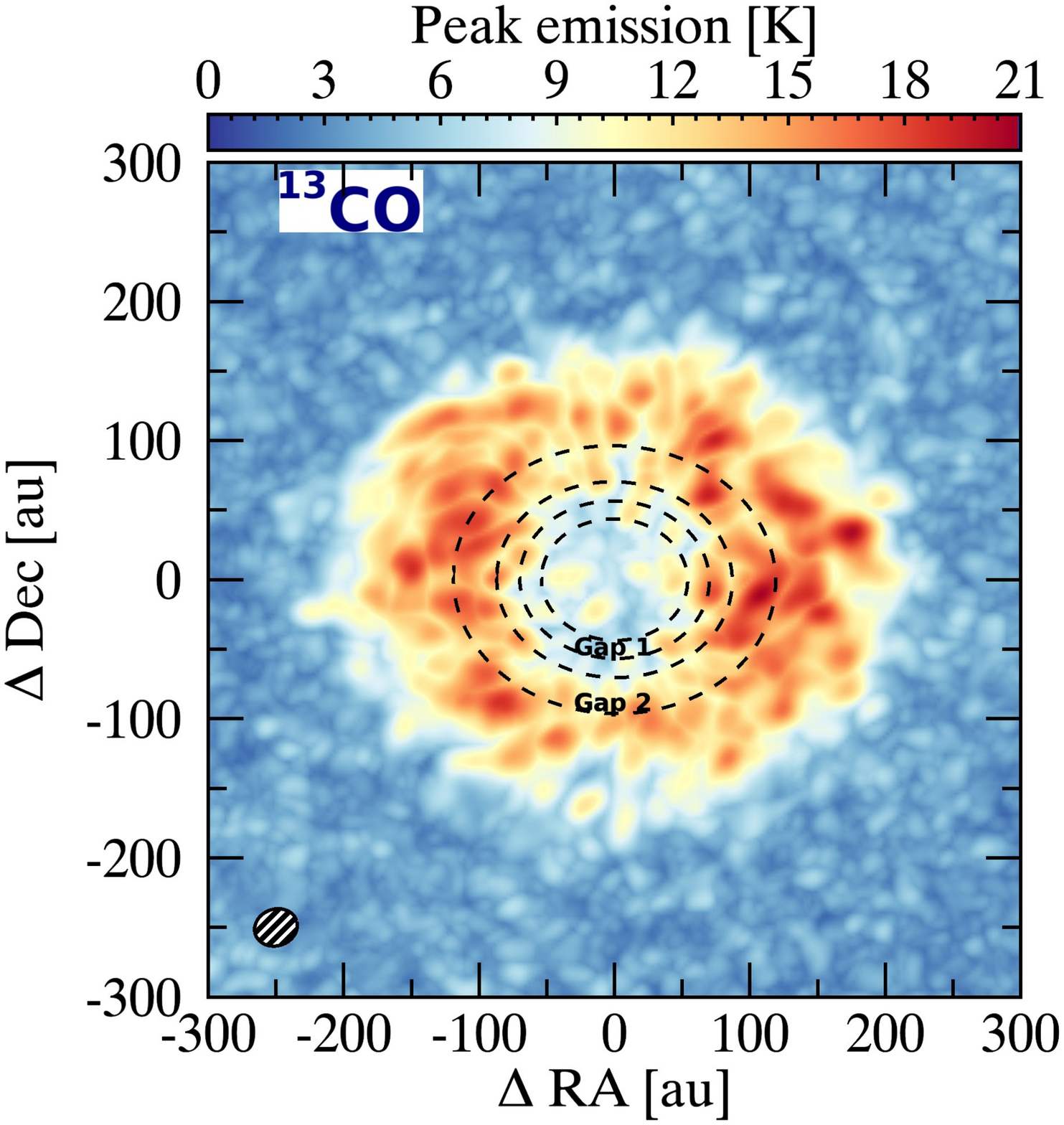}
\includegraphics[trim={0 4cm 0 4cm},angle=0,width=4.2cm]{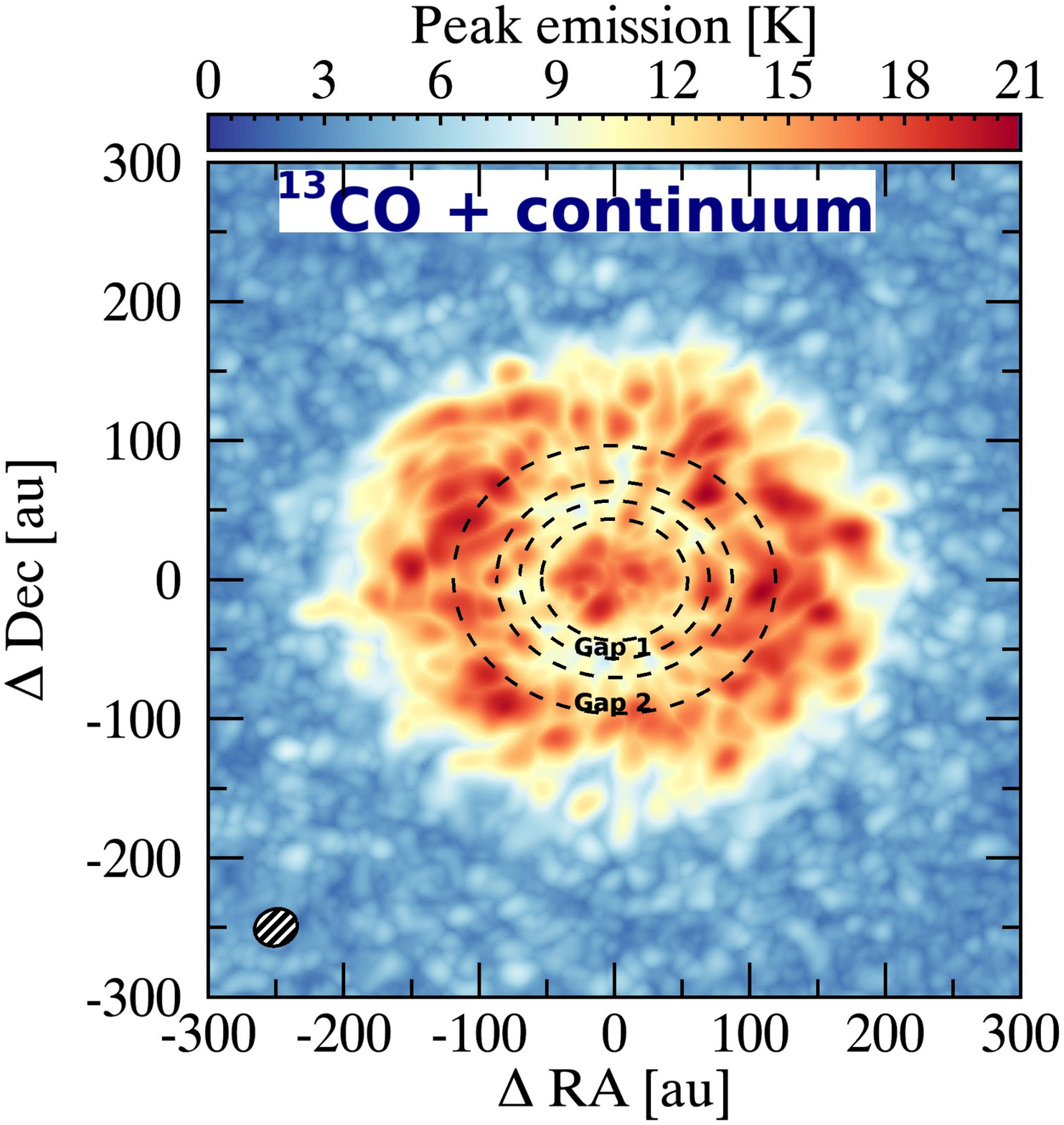}

\includegraphics[trim={0 4cm 0 4cm},angle=0,width=4.2cm]{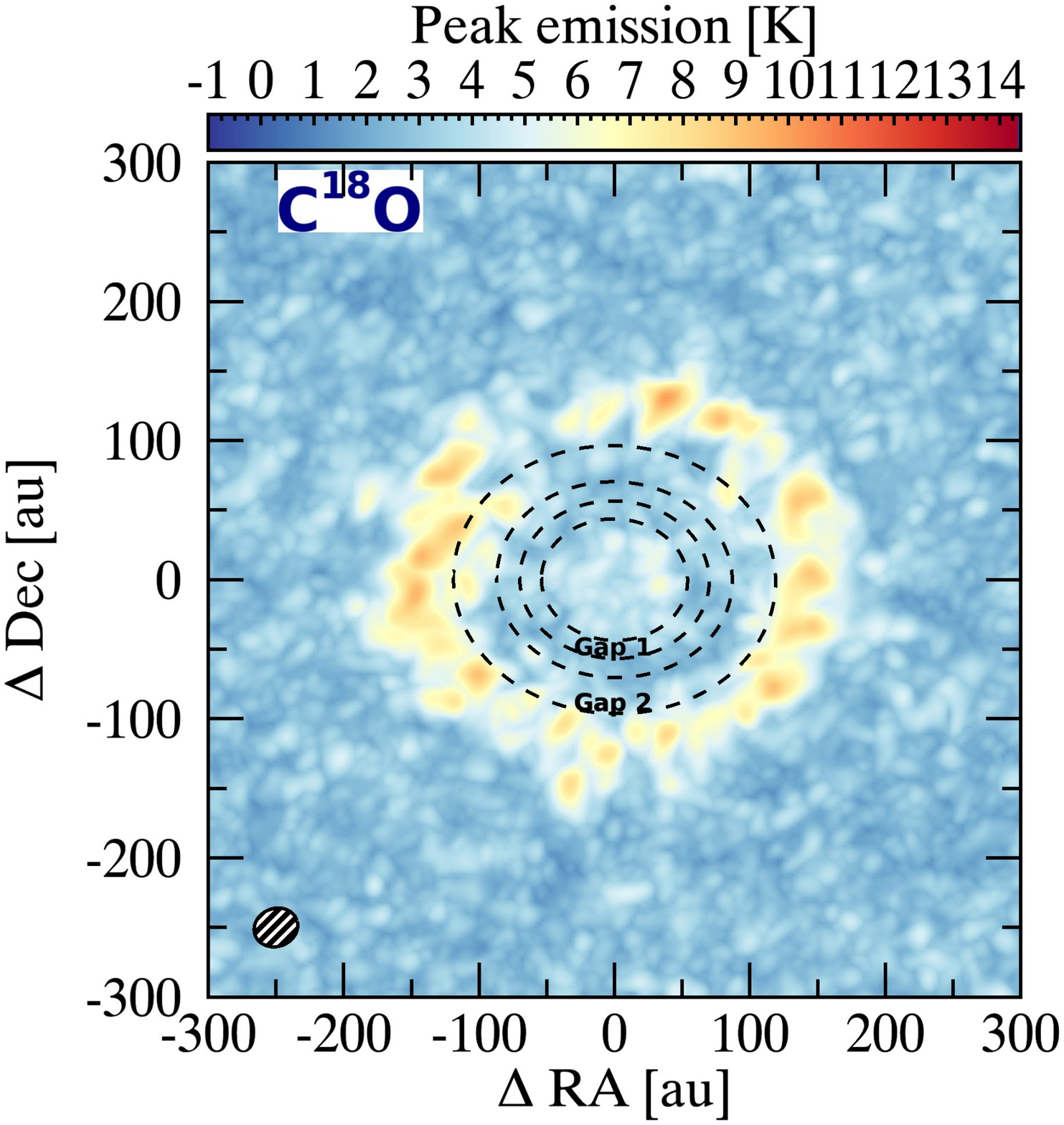}
\includegraphics[trim={0 4cm 0 4cm},angle=0,width=4.2cm]{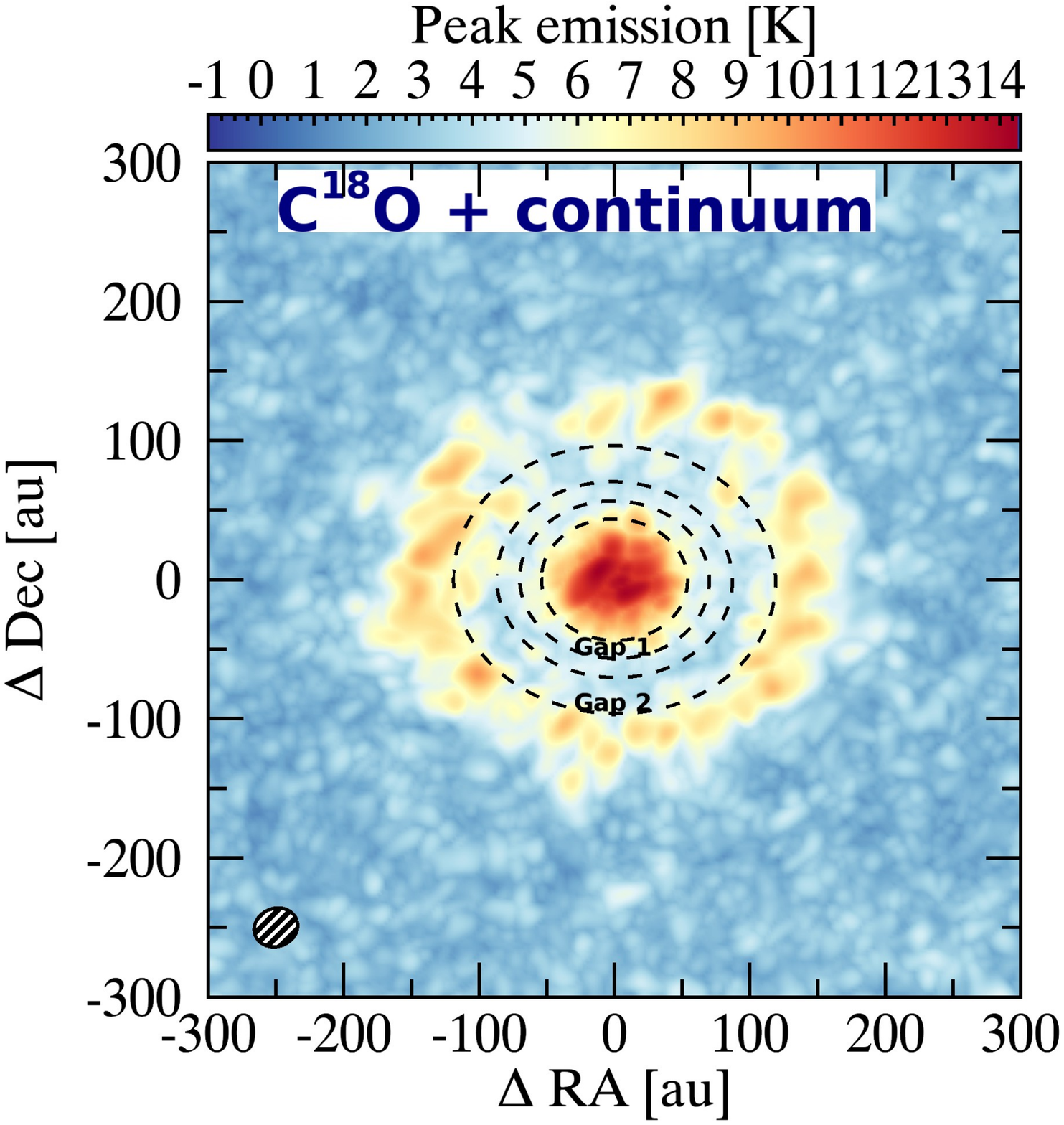}

\includegraphics[trim={0 4cm 0 4cm},angle=0,width=4.2cm]{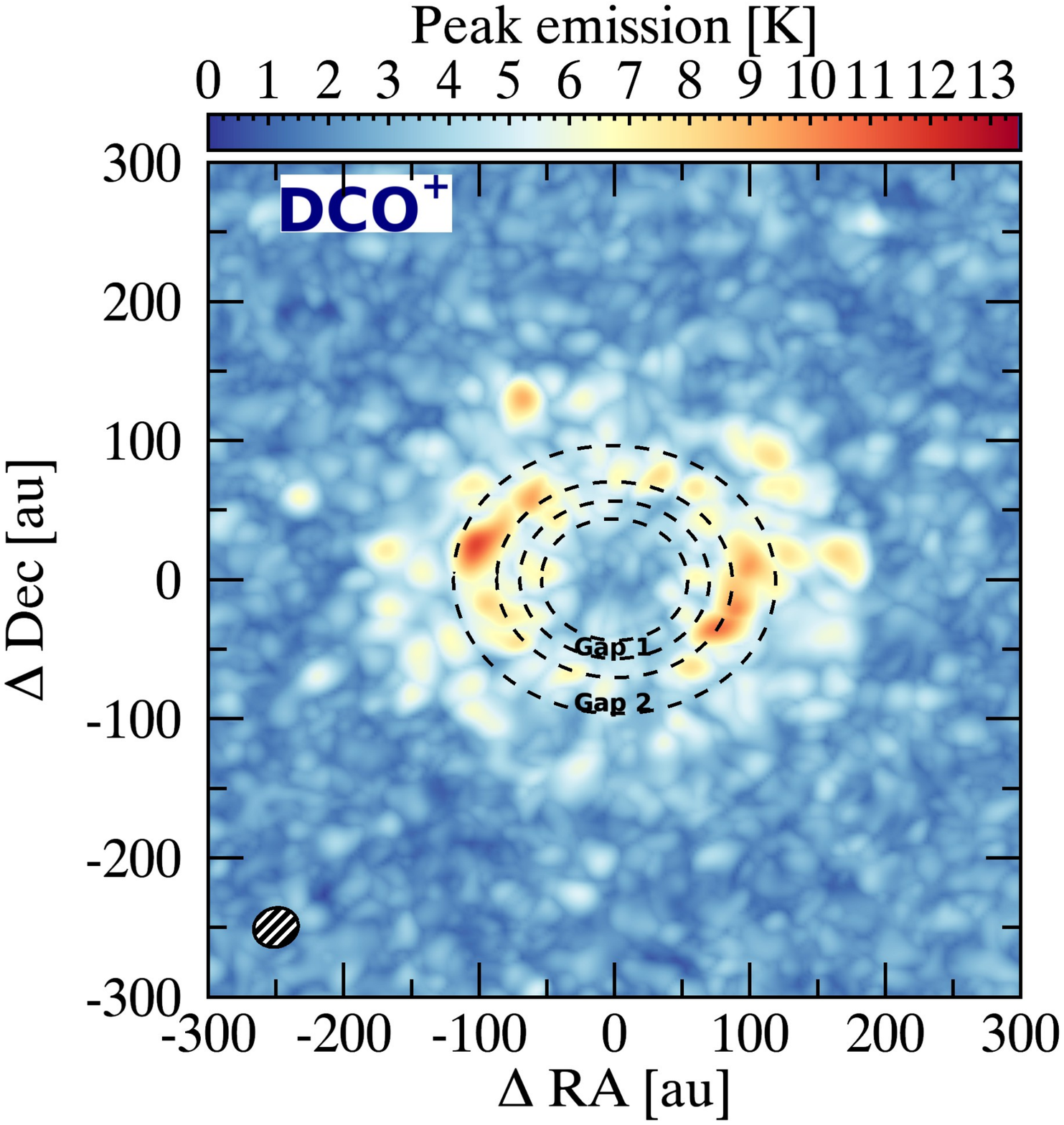}
\includegraphics[trim={0 4cm 0 4cm},angle=0,width=4.2cm]{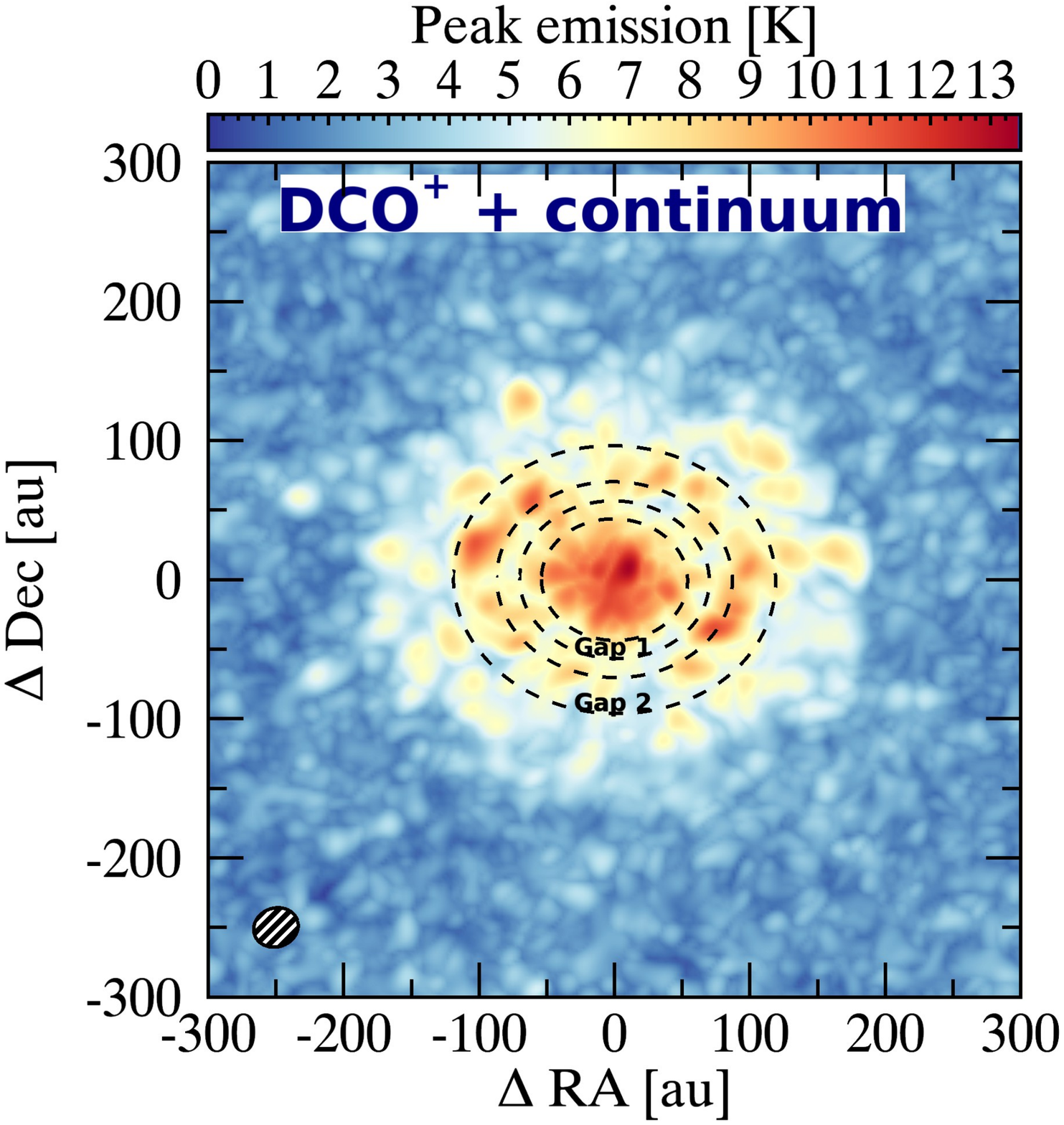}
\caption{{\it Left column:} AS~209 maximum peak intensity emission maps in units of the brightness temperature, for CO, $^{13}$CO, C$^{18}$O (2-1) and DCO$^{+}$ (3-2) emission lines from continuum subtracted data. {\it Right column:} same but from native data without continuum subtraction. All the maps are overlaid with the two continuum gaps (black dashed lines) reported in Paper~I.}
\label{fg3}
\end{figure}

These estimates are consistent with the maximum peak intensity maps \citep[moment 8, see further details on mom8 maps in][]{Boehler:2017} for the molecular emission lines that are displayed in Figure~\ref{fg3}. More specifically, Fig.~\ref{fg3} shows moment 8 maps generated with and without continuum subtraction and one can see that the emission of C$^{18}$O and DCO$^{+}$ is optically thin while this is not the case for CO and $^{13}$CO emission.

%------------------------------------------------------------------
% --- FIGURE 3 ---
%-----------------------------------------------------------------

\subsection{Radial intensity profiles}

Figure~\ref{fg1} show the azimuthally averaged 1.3~mm dust continuum radial intensity profile along with that of $^{13}$CO, CO and DCO$^{+}$ that have all been deprojected for the AS~209 disk inclination (PA=86$\degr$ and $i$= 35$\degr$, see Paper~I). Regarding CO, the displayed profile was averaged at $\pm$20$\degr$ along the major west-east emission axis after deprojecting for the disk inclination due to the source geometry \citep[see above and also][]{Teague:2018b} and in order to reduce the noise in the radial profile.
It is apparent that the CO and $^{13}$CO emission decrease with increasing radius, which is consistent with previous observations by \citet{Huang:2016}. The opacity of both lines prevents us to see further details.

One notable feature of Fig.~\ref{fg1} is that the C$^{18}$O radial intensity profile follows that of the dust continuum and harbours a drop in intensity between the two continuum gaps. Another salient result is that the DCO$^{+}$ radial intensity profile is clearly anticorrelated with that of C$^{18}$O and continuum: while the DCO$^{+}$ intensity increases or decreases with increasing radius, that of C$^{18}$O shows the opposite trend. In particular, the DCO$^{+}$ surface density is seen between the two continuum gaps.

\begin{figure}
\centering
% trim={<left> <lower> <right> <upper>}
\includegraphics[trim={1.8cm 2cm 1.4cm 2cm},angle=0,width=09cm]{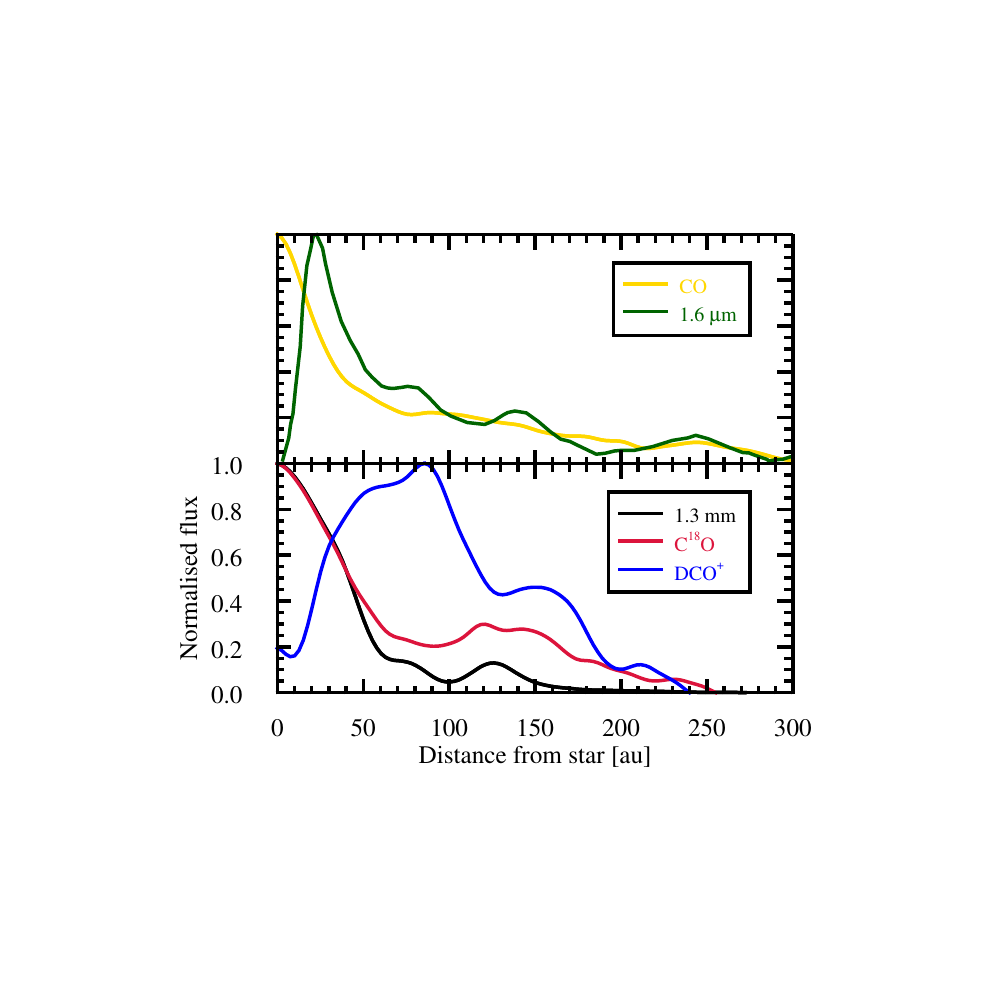}\\
\caption{ {\it Top Panel:} Normalized radial intensity profile of the CO (yellow) continuum-subtracted emission. The normalized, and scale with the radius$^{2}$, azimuthally averaged surface brightness of the 1.6~$\mu$m scattered light \citep[VLT/SPHERE H-band, adapted from][]{Avenhaus:2018} is displayed in green color. {\it Bottom Panel:}
Normalized radial intensity profiles of the continuum emission at 216~GHz (black) along with that of the C$^{18}$O (red) and DCO$^{+}$ (blue) continuum-subtracted emission (see also Fig.~\ref{fg1}).}
 \label{fg4}
\end{figure}

Finally, Fig.~\ref{fg4} displays the AS~209 CO, C$^{18}$O and DCO$^{+}$ radial profiles along with that of the 1.3~mm continuum emission and the 1.6$\mu$m scattered light radial profile as observed in H-band with VLT/SPHERE \citep[see][]{Avenhaus:2018}.  As noted by \citet{Teague:2018b}, the radial location of both the dust rings and emission peaks are offset between the two grains population, and CO emission peak at about 250~au coincide with that of the 1.6$\mu$m scattered light. Incidentally, it is interesting to note that both molecular and $\mu$m scattered light extend beyond the mm continuum emission.  

%------------------------------------------------------------------
% --- FIGURE 4--
%-----------------------------------------------------------------

%-----------------------------------------------------------------
%-----  Modeling-----------
%-----------------------------------------------------------------
\section{Gas gaps thermochemical Modelling}
\label{modeling}

In this section we aim to investigate whether the deficit of the C$^{18}$O emission is due to perturbations of the gas surface density. As we aim not to perfectly fit the disk temperature and density structures but rather to analyse how gas gaps can affect the molecular emission, we have only performed a simple parametric study of the gas perturbation as described below.

\subsection{Model description}

We used the thermochemical disk model {\it Dust And LInes} \citep[DALI, see][for a complete description]{Bruderer:2012,Bruderer:2013}, which includes the following updates. {(i)} CO isotope-selective photodissociation \citep{Miotello:2014,Miotello:2016}, to reproduce the CO isotopologues emission maps shown in Fig~\ref{fg1}. {(ii)} Density scaling factors to reproduce the AS~209 dust and gas gaps, as described below.

%-----------------------------------------------------------------------------------------------
% --- TABLE 2 ---
%-----------------------------------------------------------------------------------------------
\begin{table*}
\begin{center}
\caption{\label{tab2}Adopted DALI parameters.}
\begin{tabular}{lll} 
\hline \hline
Fixed parameter & Value  & Description\\
\hline
$h_{c}$ & 0.133 & Scale height  \\ 
$\psi$  & 0.1 & Flaring exponent\\
$i$ [$\degr$] &35\tablenotemark{a} & Disk inclination \\
$PA$ [$\degr$]  &86\tablenotemark{a}& Disk position angle \\
d [pc] & 126\tablenotemark{a,b} & distance of the source \\
M [M$_{\odot}$] & 0.9\tablenotemark{a,c} & Stellar mass\\
L [L$_{\odot}$] &1.5\tablenotemark{a,c}&Stellar luminosity\\
M$_{gas}$ [M$_{\odot}$] & 3.0$\times10^{-3}$ & Disk gas mass\tablenotemark{d}\\
M$_{dust}$ [M$_{\odot}$] &  3.5$\times10^{-4}$ & Disk dust mass\\
\hline
\hline
Adopted parameter &Values\tablenotemark{e}   & Description\\
\hline
$\gamma_{gas}$ &0.1, {\bf 0.2}, 0.3, 0.4, 0.5& $\Sigma_{gas}$ exponent \\
R$_{\rm c,gas}$ [au] & 80, 100, {\bf 120}, 140, 160 & $\Sigma_{gas}$ cutoff radius\\ 
$\delta_{gap}$& {\bf 0.1}, 0.2, 0.3, 0.4  & $\Sigma_{gas}$ scaling factor for the gaps\\ 
 $\delta_{ring}$ & {\bf 0.4}, 0.6, 0.8, 1.0 & $\Sigma_{gas}$ scaling factor for the ring\\ 
\hline
\end{tabular}
\end{center}
\tablenotetext{a}{Paper~I.}
\tablenotetext{b}{\citet{Gaia-Collaboration:2016}.}
\tablenotetext{c}{\citet{Andrews:2009}.}
\tablenotetext{d}{The following standard elemental abundances of carbon and oxygen of 1.35$\times$10$^{-4}$ and 2.88$\times$10$^{-4}$ are assumed.}
\tablenotetext{e}{The adopted values are given in boldface.}
\tablecomments{We refer to Paper~I for the remaining parameters.}
\end{table*}
%++++++++++++++++++++++++++++++++++++++++++++++++++++++++++++++++++

The starting point is the best-fit model of the dust presented in Paper~I, which reproduces the 1.3~mm continuum emission. For the dust surface density profile we used that of Paper I. For simplicity, the surface density profile of the small grains population is assumed to follow that of the large grains (for further details see Section 4.1 of Paper~I). The gas density surface density, $\Sigma_{\rm gas}$, is defined as follows :
\begin{equation}\label{eq:dali_sd}
\Sigma_{\rm gas}(R) = \delta (R) \Sigma_{\rm c,gas}\Bigg(\frac{R}{R_{\rm c, gas}}\Bigg)^{\gamma_{gas}}\exp\Bigg[ - \Bigg( \frac{R}{R_{\rm c,gas}} \Bigg)^{2-\gamma_{gas}} \Bigg],
\end{equation}
where $R_{c,gas}$ is the cutoff radius of the gas, $\gamma_{gas}$ is the surface density power-law exponent and  $\delta (R)$ is the surface density scaling factor within the gaps described as follows (by definition $\delta (R)$ $>$ 0):

\begin{eqnarray}
\delta(R) =\left\{
\begin{array}{ll}
\delta_{gap}  & \text{for R $\in$ [54\,au, 70\,au] $\&$ [87\,au, 119\,au]}\\
\delta_{ring}  & \text{for R $\in$ [70\,au, 87\,au]}\\
1 & \text{otherwise.}
\end{array}
           \right.
           \end{eqnarray}
We note that owing to the data angular resolution (allowing us to probe only scale $\ge$ 30~au), we assumed the same $\delta$(R) in the two gaps.

In order to reproduce both the absolute flux level and the radial profile of the three CO isotopologues ($J$=2-1) transitions, we first run a grid of DALI models varying $\Sigma_{\rm c,gas}$\footnote{For the total gas mass, we assumed the following standard elemental abundances of carbon and oxygen of 1.35$\times$10$^{-4}$ and 2.88$\times$10$^{-4}$, respectively.}, $\gamma_{gas}$ and R$_{\rm c,gas}$, while keeping fixed the scale height\footnote{The scale height distribution is described by: $  h = h_{c}(\frac{R}{R_c})^{\psi} $ (for further details see Paper I).} $ h_{c}$ (0.133) and the flaring exponent $\psi$ (0.10) as in paper I. Table~\ref{tab2} lists the DALI parameter values adopted in the present study.

In a second step, a grid of DALI disk structures has been created by varying $\delta_{gap}$ and $\delta_{ring}$ (the values are given in Table~\ref{tab2}). The data-model comparison is performed in the image plane: DALI creates synthetic channel maps which are then convolved with a Gaussian beam of the same size as the beam of the observations (see Table~\ref{tab1}). Then, the latter are collapsed to create integrated intensity maps.

%------------------------------------------------------------------
% --- FIGURE 5---
%-----------------------------------------------------------------
\begin{figure*}
\centering
% trim={<left> <lower> <right> <upper>}
\includegraphics[angle=0,width=18cm]{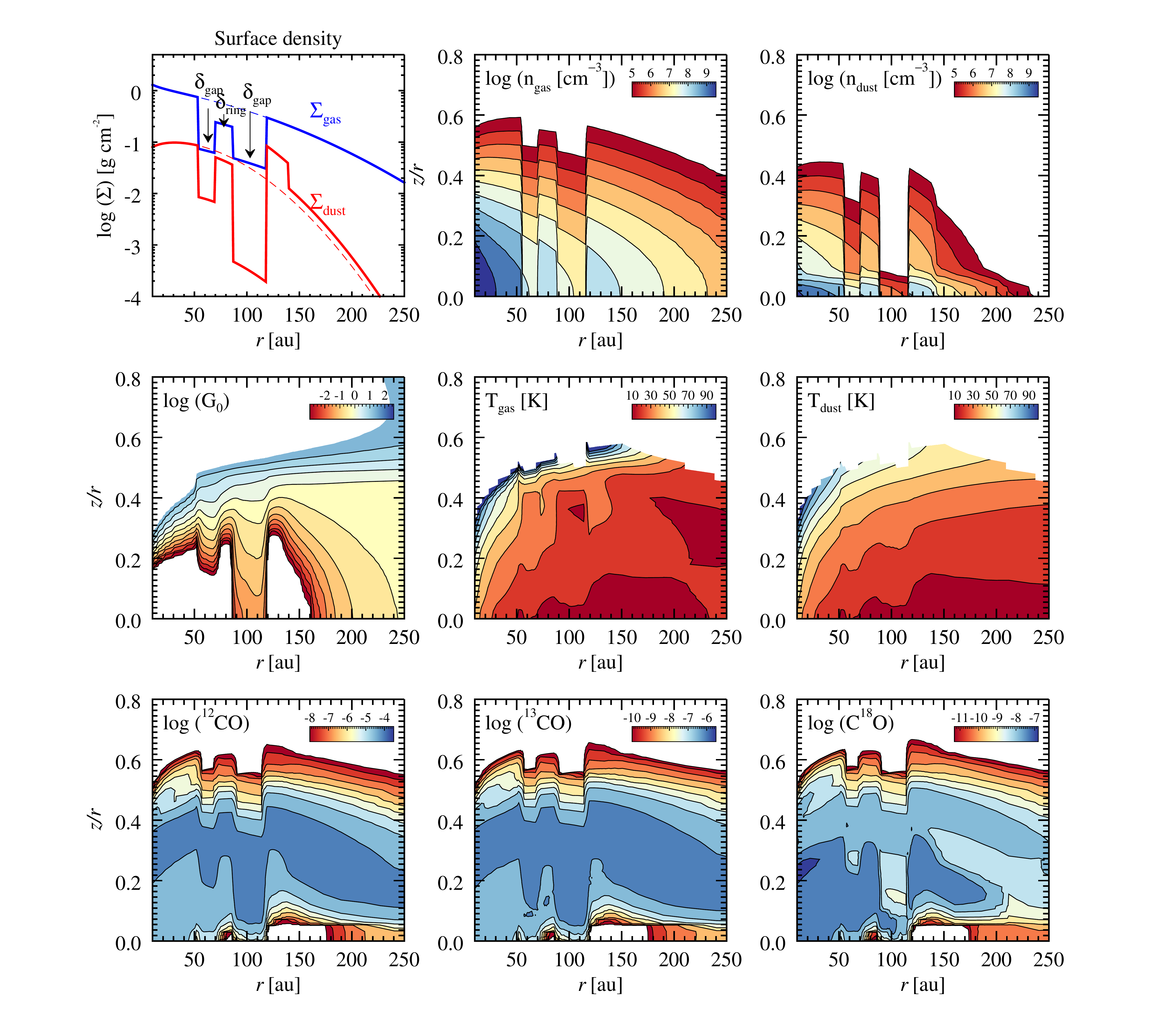}
\caption{{\it Top row:} Surface density parametrisation and gas and dust volume densities used as input of a random modelling of gas gaps with DALI code for AS~209. {\it Middle and bottom rows:} Output of the thermochemistry: FUV radiation field G$_0$, gas and dust temperatures, and $^{12}$CO, $^{13}$CO and C$^{18}$O abundances.  Where not specified, the y-axis refers to the vertical scale-heights (z/r).}
\label{fg5}
\end{figure*}

\subsection{Modelling results}

\begin{figure*}
\centering
% trim={<left> <lower> <right> <upper>}
\includegraphics[angle=0,width=4.4cm]{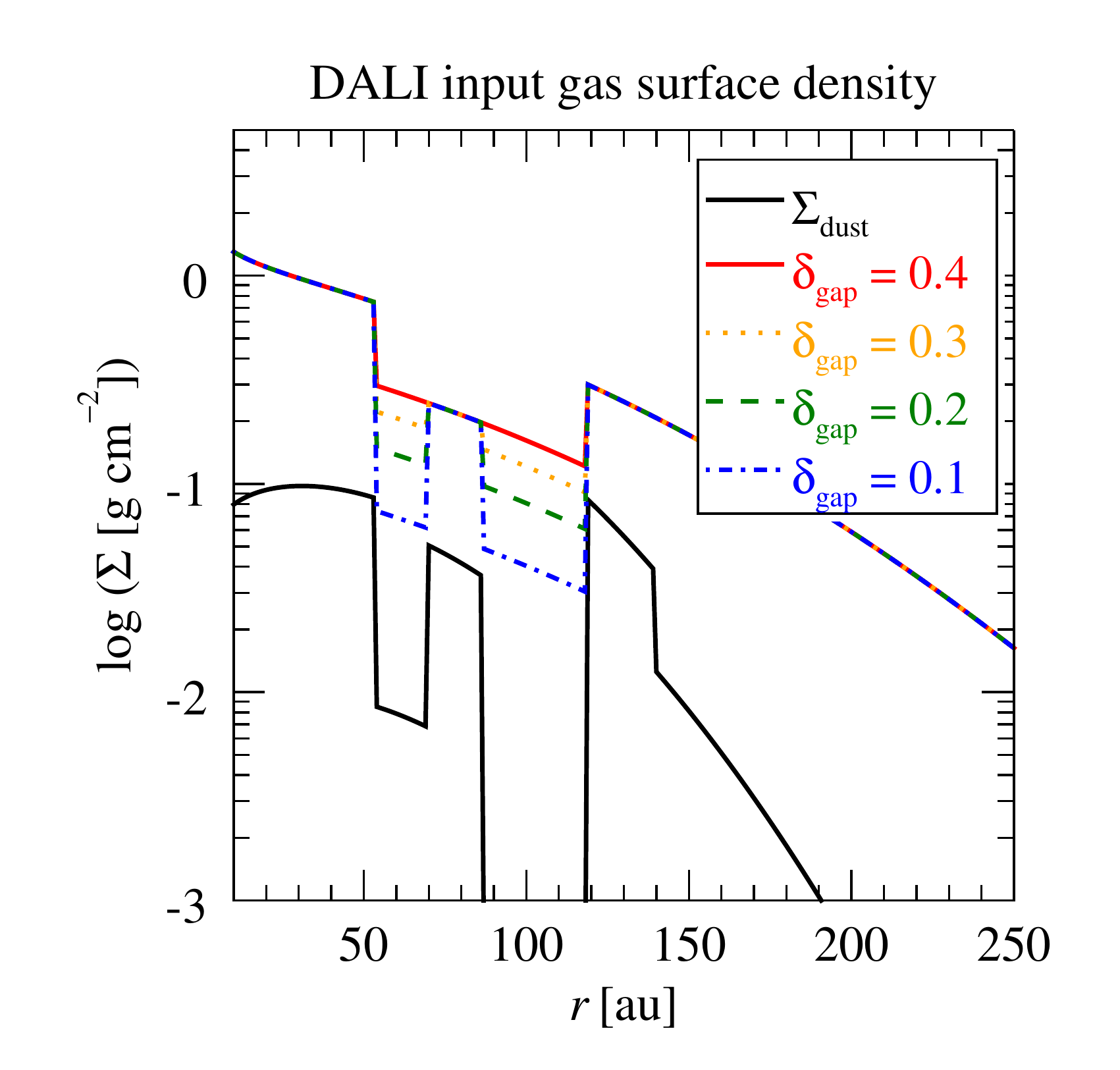}
\includegraphics[angle=0,width=4.4cm]{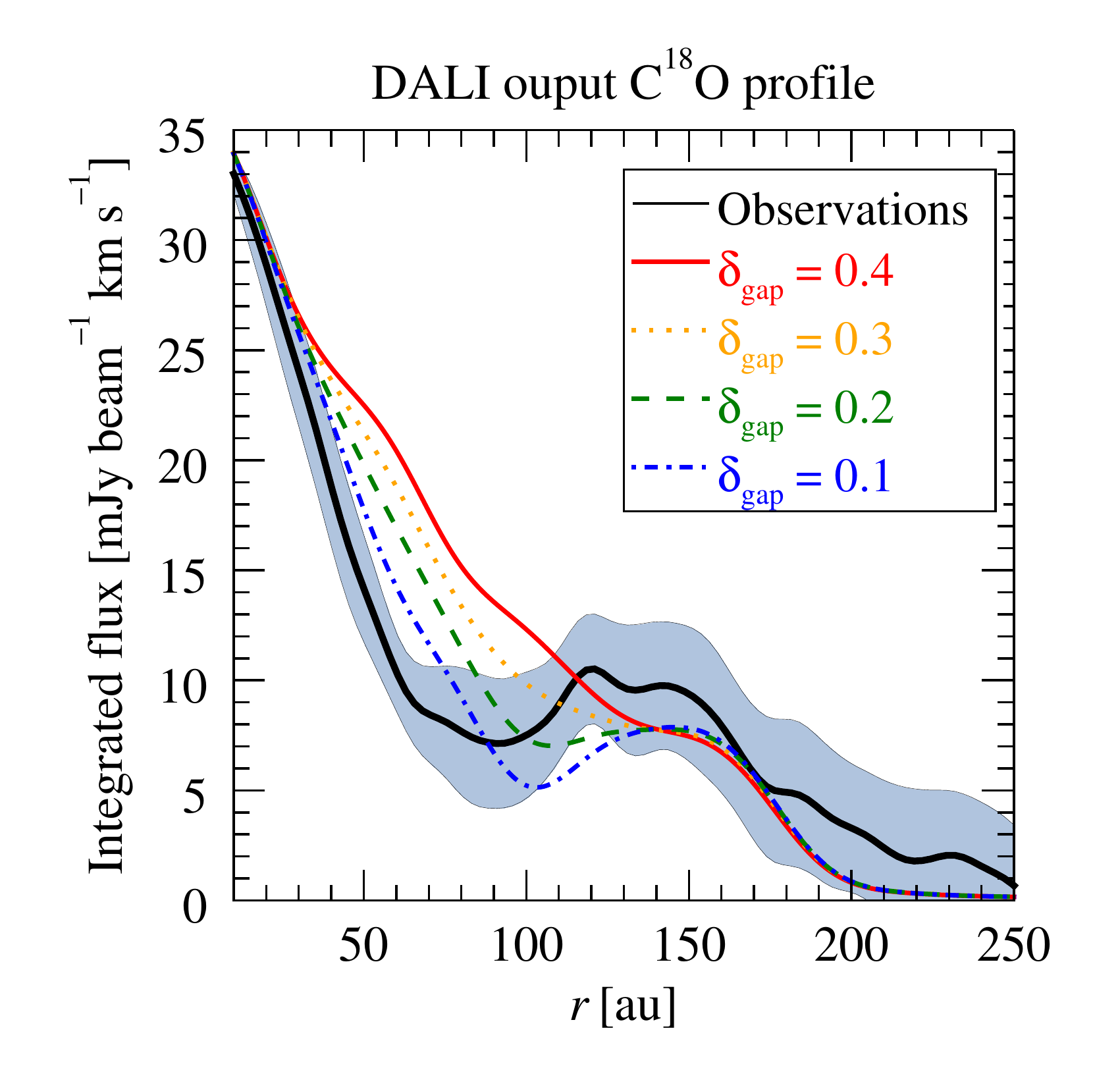}
\includegraphics[angle=0,width=4.4cm]{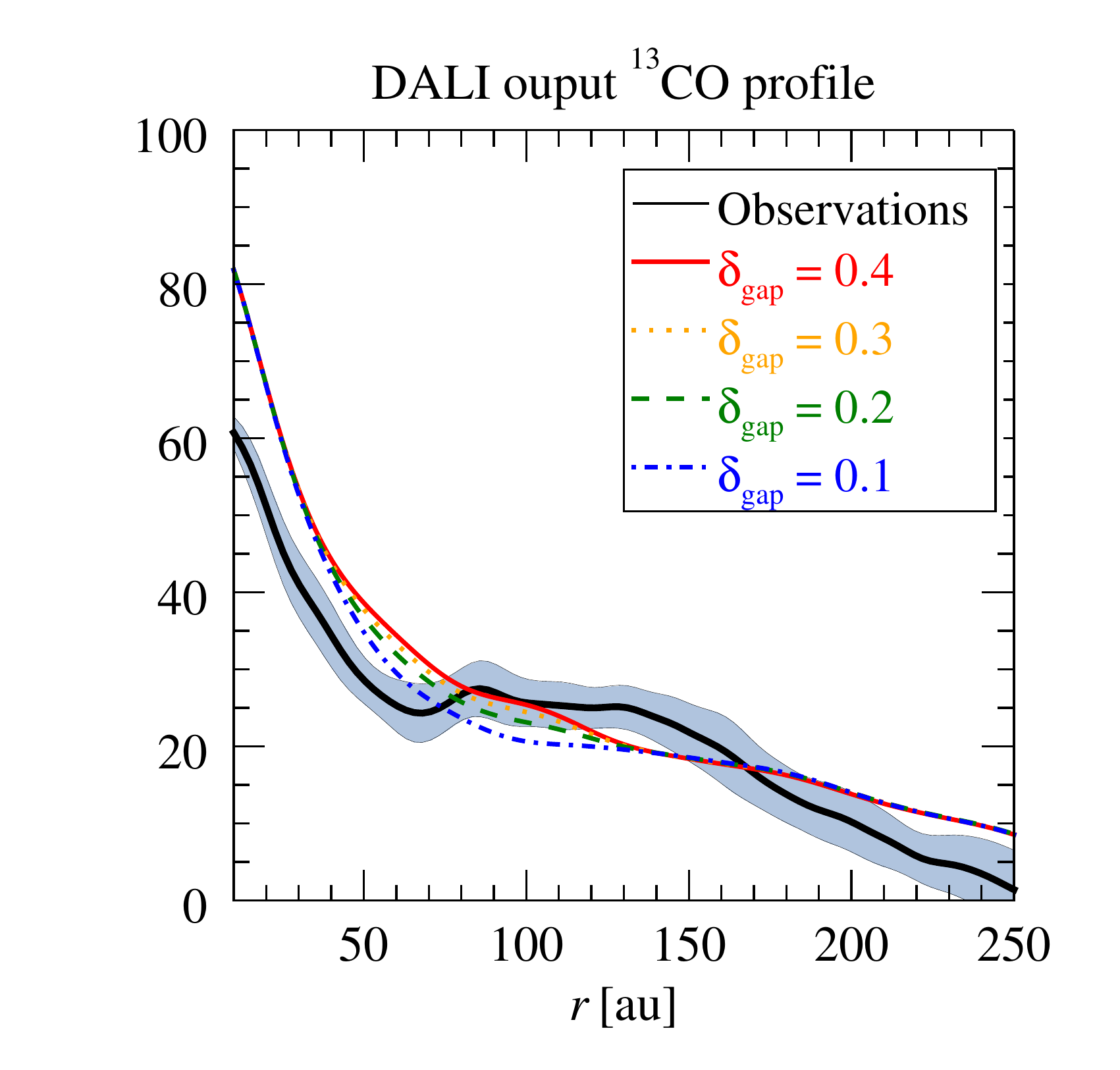}
\includegraphics[angle=0,width=4.4cm]{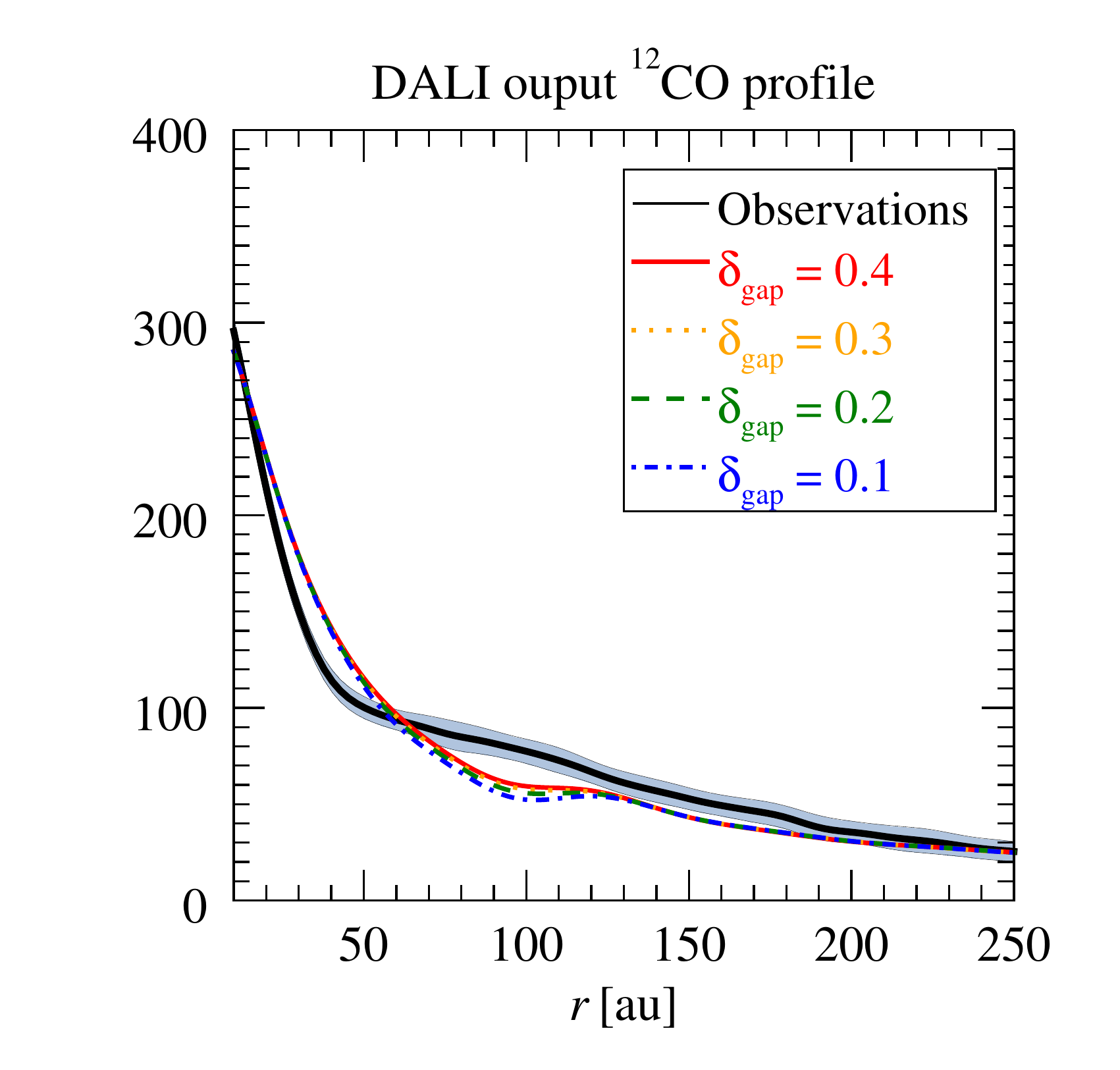}
\includegraphics[angle=0,width=4.4cm]{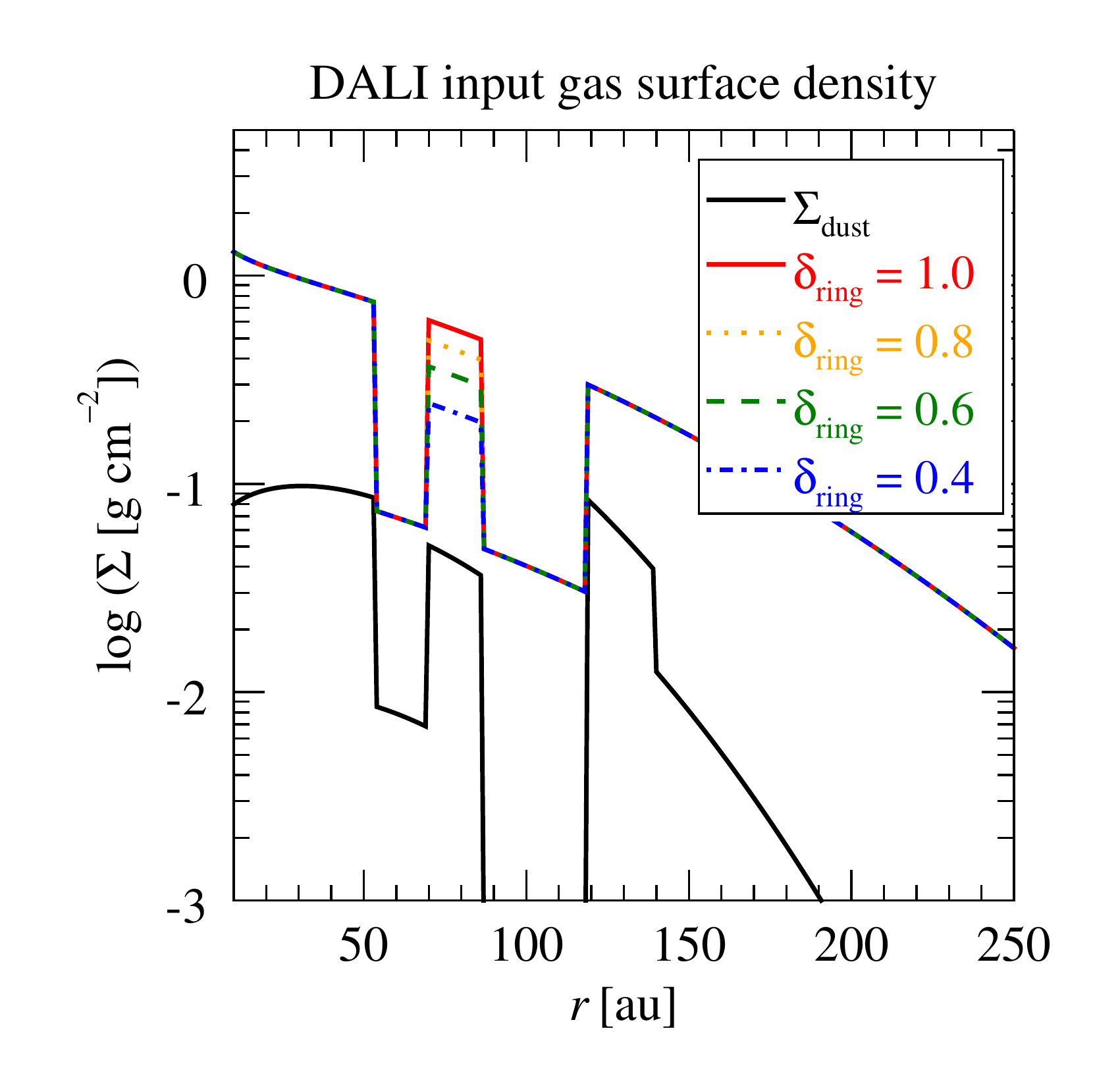}
\includegraphics[angle=0,width=4.4cm]{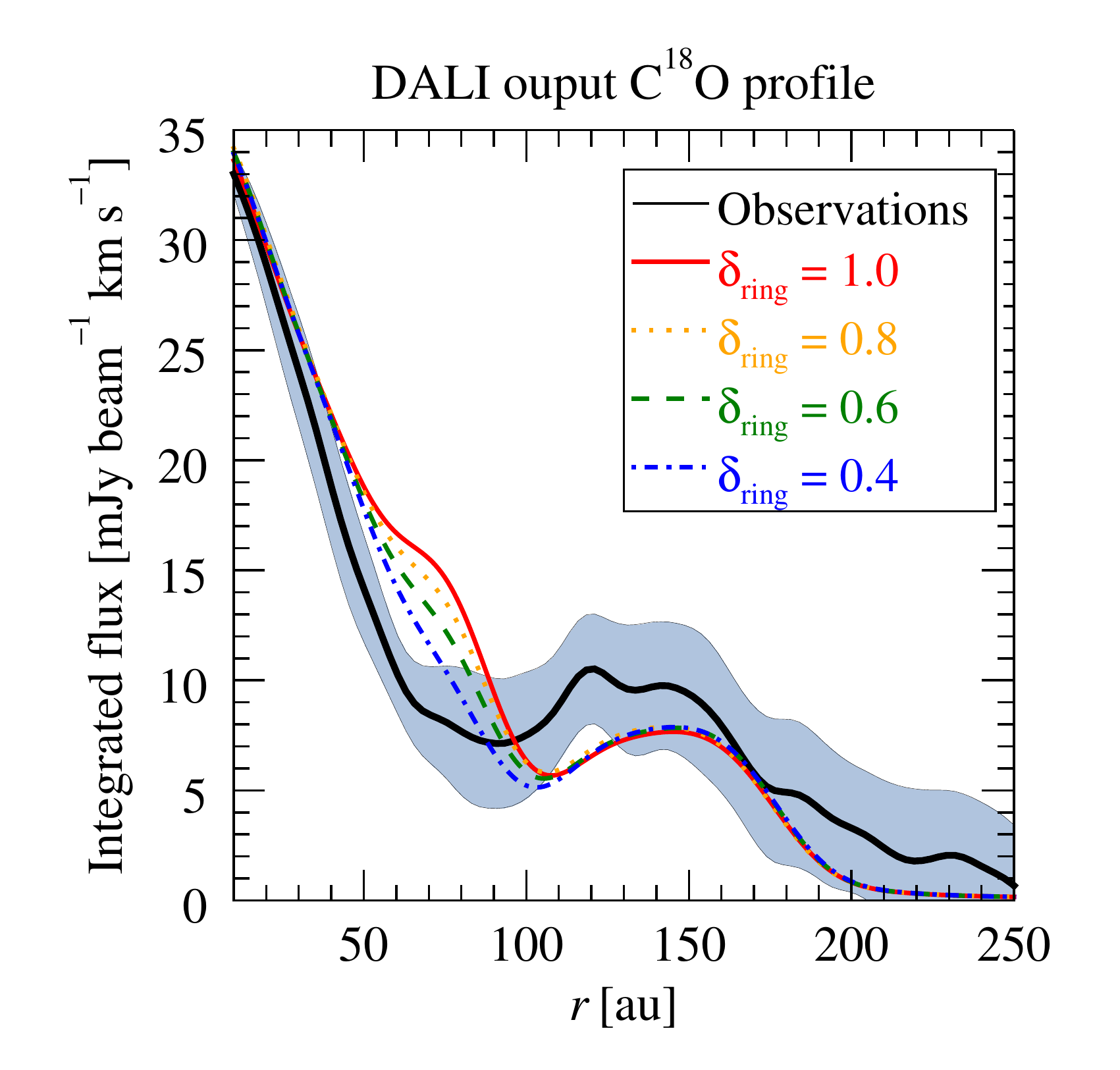}
\includegraphics[angle=0,width=4.4cm]{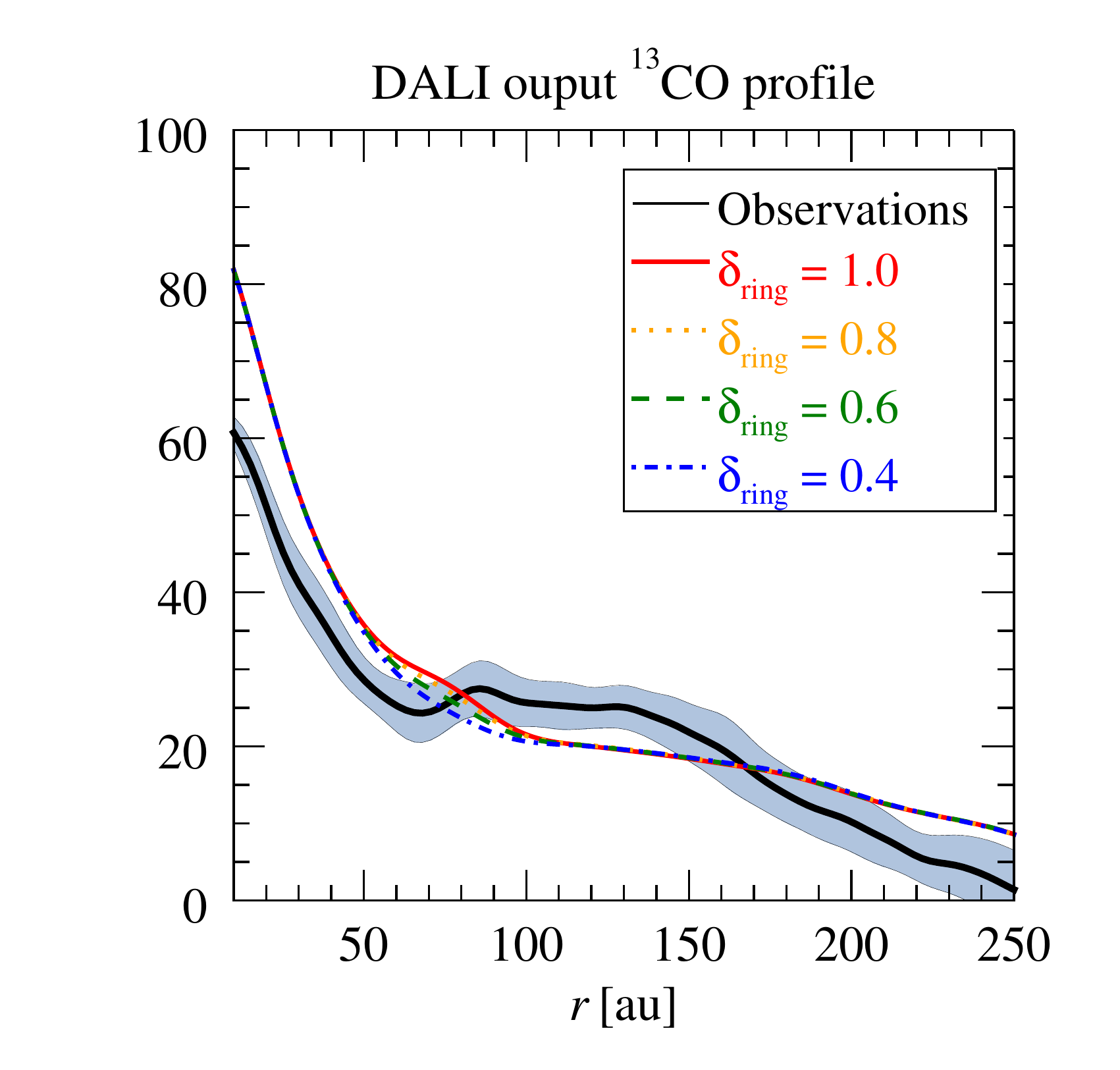}
\includegraphics[angle=0,width=4.4cm]{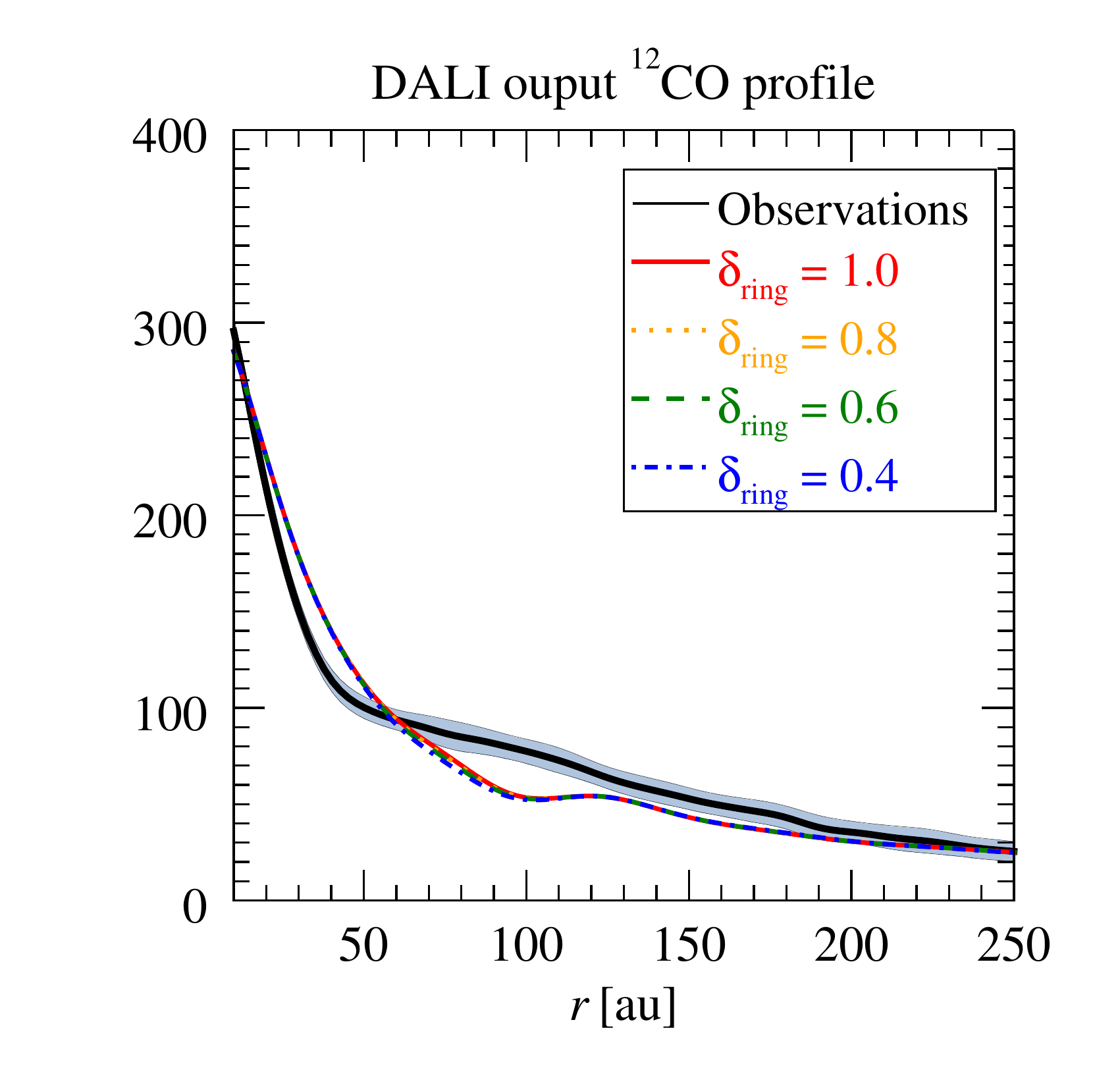}
\caption{{\it Left column:}  Surface density profiles used as input for DALI modelling. That of the dust is shown in black while those of the gas are shown in different line styles and colours, each referring to different depletion factors for $\delta_{gap}$ and $\delta_{ring}$. {\it Other columns:} Output of DALI models showing the resulted predicted C$^{18}$O, $^{13}$CO and $^{12}$CO radial density profiles, respectively overlaid with the observed one (see Fig.~\ref{fg1}). We note that the modeled $^{13}$CO profile is scaled down by 20$\%$ to match the absolute flux level. Where not specified $\delta_{\rm gap}$=0.1 and $\delta_{\rm ring}$=0.4.}
\label{fg6}
\end{figure*}

Figure~\ref{fg5} shows the surface density parametrisation along with the density and temperature structures of a random DALI model and the resulting abundance of the three CO isotopologues in presence of gas gaps. Figure~\ref{fg6} shows the input surface density structure for the different values of $\delta_{\rm gap}$ and $\delta_{\rm ring}$  along with the resulting deprojected and azimuthally averaged along with the observed CO, $^{13}$CO and C$^{18}$O radial profiles. The adopted gas density structures reproduces well the overall $^{13}$CO and $^{12}$CO profiles although minor differences are seen. The latter are mainly due to the uncertainties in the vertical density and temperature structures \citep[see][]{Bruderer:2012}.

Our modelling clearly shows that in order to reproduce the observed C$^{18}$O deficit, $\Sigma_{\rm gas}$ needs to be lowered, in the gaps, by applying a scaling factor of about  0.1--0.2 compared to the unperturbed profile (i.e. $\delta$(R)=1). Our gas gaps thermochemical modelling also shows that the drop of C$^{18}$O is not the result of temperature and/or opacity effects alone (the emission of $J=2-1$ transition being thin) but that of an intrinsic gas density drop. Indeed, without gas gaps, our model cannot reproduce the C$^{18}$O profile (see Figure~\ref{fg6}). Interestingly enough, the $^{13}$CO and $^{12}$CO $J=2-1$ emission remains optically thick within the gaps. As a consequence, the modelled surface density perturbations lead to minor changes in the resulting $^{13}$CO and $^{12}$CO radial profiles as shown in Figure~\ref{fg6}.

%------------------------------------------------------------------
% --- FIGURE 6---
%-----------------------------------------------------------------

Finally, it is interesting to note that UV irradiation penetrates deeper in the disk layers where gap dusts are located as shown in Figure~\ref{fg5}. In that instance, the FUV radiation field G$_0$ is larger and the C$^{18}$O/$^{12}$CO mass ratio is further reduced because of photochemistry. From the measured $^{12}$CO/C$^{18}$O flux ratio, we estimate that isotope-selective photodissociation \citep{Visser:2009} contributed to about 10--20$\%$ on the observed C$^{18}$O intensity inside the gaps. 

We stress that our present modelling does not fully reproduce the observations because of uncertainties on the 2D temperature structure. A more detailed modelling will be presented in an upcoming paper (Fedele et al. in prep.), however.

%
%-----------------------------------------------------------------
%-----  Discussion-----------
%-----------------------------------------------------------------
\section{Discussion}
\label{discussion}
%-----------------------------------------------------------------

\subsection{Comparison with hydrodynamical simulations}

\begin{figure}
\centering
% trim={<left> <lower> <right> <upper>}
\includegraphics[trim={4.5cm 5.5cm 3cm 4cm},clip,angle=0,width=8cm]{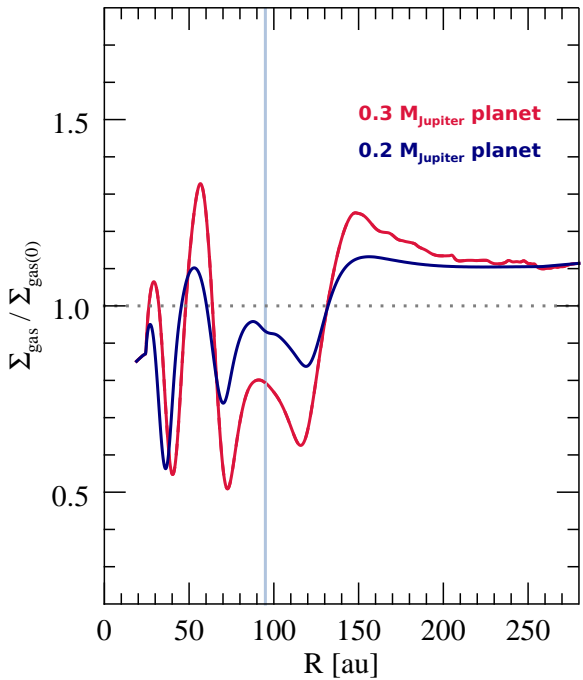}\\
\caption{Hydrodynamical simulations of the gas surface density ($\rm \Sigma_{gas}$) perturbed by the presence of a 0.2~M$\rm_{Jupiter}$ (blue) planet and a 0.3~M$\rm_{Jupiter}$ (red) planet located at 95~au. The resulting perturbations are displayed with-respect-to the unperturbed gas surface density ($\rm \Sigma_{gas(0)}$). The location of the planet is indicated by a vertical line.}
\label{fg7}
\end{figure}

From our 1.3~mm continuum observations coupled with hydrodynamical simulations, we have previously shown that the presence of a giant planet or of a pair of planets likely explains the observed dust gaps (see Paper~I).
Recently, from 2D hydrodynamical simulations of planet--disk interaction, \citet{Facchini:2018} have shown that gaps opened by a planet leads to a thermal gas and dust decoupling. The latter strongly affects the dust-to-gas mass ratio within the gaps with respect to the overall ratio. Our DALI modelling of the dust and C$^{18}$O emission is consistent with this interpretation. In addition, hydrodynamical simulations of planet-disk models by \citet{Facchini:2018} show that the presence of a giant planet not only affects the dust radial intensity profile but also the ones of CO and its isotopologues. Indeed, the latter harbour a gap \citep[whose depth and width both depend on the planet mass and disk viscosity, see e.g.][]{Lin:1986,Crida:2006,Fung:2014,Duffell:2015,Durmann:2015,Kanagawa:2017} at the planet location in the disk. In that context, we have thus performed hydrodynamical simulations to further investigate the hypothesis of a planet producing gas gaps in the AS~209 disk. More specifically, we have carried out 2D hydrodynamical simulations using FARGO-3D \citep[see further details in paper~I, and in][]{Benitez-Llambay:2016,Rosotti:2016} for a planet of 0.2~M$\rm_{Jupiter}$ and one of 0.3~M$\rm_{Jupiter}$ located at 95~au. The simulations were run for 1000 orbits of the planet (corresponding to about 10$^6$ years) and are inviscid (i.e. the viscosity, $\alpha$, is null).
 The simulations used a grid extending from 19 to 285~au, using 1766 logarithmically spaced cells in radius and 4096~cells in azimuth. Wave-killing boundaries were used in the radial direction \citep{de-Val-Borro:2006}. The disk aspect ratio was taken to be $\rm 0.1(R/100\,\mathrm{au})^{0.225}$ as in Paper~I. The results are shown in Figure~\ref{fg7}. It is immediately apparent that the presence of a planet perturbes the gas surface density by inducing a gap which is wider and deeper for a massive planet. This is commensurate with our C$^{18}$O observations and the results of Section~\ref{modeling}. We infer that the optical thickness of the $^{13}$CO and CO line prevents us to see any gap in their radial distribution profiles as they only probe the disk surface \citep[see Section~\ref{modeling} and][]{Ober:2015}. Incidentally, the hydrodynamical simulations show a bump at the location of the planet due to the so-called co-rotation zone. This feature is not seen in our data because we do not have sufficient sensitivity.

%------------------------------------------------------------------
% --- FIGURE 7 ---
%-----------------------------------------------------------------

\subsection{On the physical nature of the $\rm C^{18}O$ gap}
Our modelling of the CO isotopologues emission shows that the drop of C$^{18}$O between the two dust gaps is the result of an intrinsic gas density drop (see Fig.~\ref{fg6} and Section~\ref{modeling}) ; which is commensurate with our hydrodynamical simulations. Incidentally, \citet{Huang:2016} argued that the observed C$^{18}$O outer ring was due to external photodesorption processes \citep{Facchini:2016}. However, the lower angular resolution and sensitivity of their data were not sufficient enough to allow the authors to see the same structure in the dust emission. Our more sensitive combined high resolutions data clearly show that the outer gas ring seen in C$^{18}$O is also seen in the 1.3~mm dust emission (see Figs.~\ref{fg1}, \ref{fg4} and Paper I) and, therefore, rule out the hypothesis of external photodesorption. 

Our findings strongly support the scenario proposed in Paper I: the formation of  a giant planet (at least M$\rm_{planet}$$>$0.2~M$\rm_{Jupiter}$) is likely occurring in the AS~209 disk as: \textit{i)} the C$^{18}$O outer ring at 120~au is also seen in the 1.3~mm  continuum emission, \textit{ii)} the observations can only be reproduced by lowering the gas surface density and, \textit{iii)} the wide C$^{18}$O deficiency between the two continuum gaps along with the  C$^{18}$O increased beyond the dust gaps are both consistent with hydrodynamical simulations of massive planet(s) in formation. Moreover, among the proposed gap-opening mechanisms, magneto-rotational instability can only reduce $\Sigma_{\rm gas}$ by a factor of a few at the edge of the dead zone \citep{Flock:2015}, while only a planet-induced dynamical clearing is able to generate the deep gas gaps observed here.
Finally, our present data does not allow us to distinguish between the single planet scenario that can open multiple gaps for very low viscosity from the one suggesting the presence of two planets \citep[see Paper I and][]{Dong:2015,Dong:2017,Bae:2017}.

\subsection{DCO$^{+}$ a tracer of forming-planet(s)? }
One notable feature of the DCO$^{+}$ emission towards AS209 is that it harbors a ring lying in the region located between the two dust continuum gaps (see Fig.~\ref{fg1}). Interestingly enough, within these dust gaps, one can observe an overdensity of DCO$^{+}$ while neutral gas (i.e C$^{18}$O) and dust are depleted (see Figs~\ref{fg1} and \ref{fg4}).

Owing to the low density of both gas and dust within the dust gaps, the medium is likely more ionized than the surrounding regions. Indeed, ionisation processes are more efficient in regions of the disk depleted of material as the gas is less shielded by the dust. Consequently, DCO$^{+}$ enhancement  might be the result of multiple chemical pathways such as \textit{i)} deuterium exchange between CO and H$_2$D$^{+}$ at low temperatures \citep[$\le$20K,][]{Watson:1976,Pagani:1992}, \textit{ii)} that with HCO$^{+}$ and D atoms \citep{Aikawa:2018} and, \textit{iii)} with CH$_2$D$^{+}$ and CH$_4$D$^{+}$ at warmer temperatures \citep[30~K$\le$T$\le$70~K][]{Favre:2015,Carney:2018}, although owing to the distance from the central star and the disk self-shielding in the gap, this region is probably cold. A full physico-chemical modelling of the DCO$^{+}$ chemistry is needed to investigate furthermore the production of DCO$^{+}$ and will be the subject on an upcoming paper. 
Nonetheless, our finding leads us to suggest that ions enrichment at the location of neutral gas and dust deficit in protoplanetary disk could be used as proxy for planet(s) formation.

%===================================================================================================
%===================================================================================================

%----------------------------------------------------------------
%-----  Conclusions---------
%----------------------------------------------------------------
\section{Conclusions}
\label{conclusions}
Our study shows that the optically thin C$^{18}$O (2--1) and DCO$^{+}$ (3-2) emission harbors rings and displays anti-correlated radial profiles. More specifically, between the previously observed two dust gaps, there is a deficit of C$^{18}$O while DCO$^{+}$ (3-2) is enhanced. Our thermohemical modelling of CO, $^{13}$CO and C$^{18}$O implies a reduced gas surface density in correspondance of the dust gaps identified in Paper~I. These findings led us to infer that the  formation of a planet of 0.2--0.3~M$_{Jupiter}$ at about 100~au is occurring in the AS~209 protoplanetary disk. The lower limit of 0.2~M$_{Jupiter}$ is needed to reproduce the observed C$^{18}$O deficit between 60~au up to 110~au. The upper limit of 0.3~M$_{Jupiter}$ is determined by the size of the dust gaps as measured in Paper~I. 

An interesting outcome of our thermochemical modelling is that C$^{18}$O being more sensitive to the gas and dust density perturbations (in comparison to the optically thicker CO isotopologues), it is best suited to probe the presence of forming-planet(s) down to $\lesssim$ M$_{Saturn}$ in the outer disk.
 
%===================================================================================================
%===================================================================================================
%
     
%-----------------------------------------------------------------
%---------ACKNOWLEDGEMENTS-----
%-----------------------------------------------------------------
%
%\section{Acknowledgements}
\acknowledgments
We thank the referee, Dr Takayuki Muto, for his very fruitful comments that have strengthened our paper. CF acknowledges Franck Hersant for fruitful discussion on disk gaps. CF and DF acknowledge financial support provided by the Italian Ministry of Education, Universities and Research, project SIR (RBSI14ZRHR). AM acknowledges an ESO Fellowship. M.T. has been supported by the DISCSIM project, grant agreement 341137 funded by the European Research Council under ERC-2013-ADG. D.S. acknowledges support from the Heidelberg Institute of Theoretical Studies for the project ``Chemical kinetics models and visualization tools: Bridging biology and astronomy".
The PI acknowledges assistance from Allegro and Bologna, the European ALMA Regional Center nodes in the Netherlands and Italia, respectively. CF and DF thank J. Huang for sharing the ALMA-Cycle 2 data. This study makes use of ALMA data. ALMA is a partnership of ESO (representing its member states), NSF (USA) and NINS (Japan), together with NRC (Canada), NSC and ASIAA (Taiwan), and KASI (Republic of Korea), in cooperation with the Republic of Chile. The Joint ALMA Observatory is operated by ESO, AUI/NRAO and NAOJ.

%===================================================================================================
%============================= APPENDIX=======================================
%===================================================================================================

\appendix

\section{Channel emission maps}

\begin{figure}[h!]
\centering
% trim={<left> <lower> <right> <upper>}
\includegraphics[trim={3cm 4.5cm 9cm 1cm},angle=0,width=16cm]{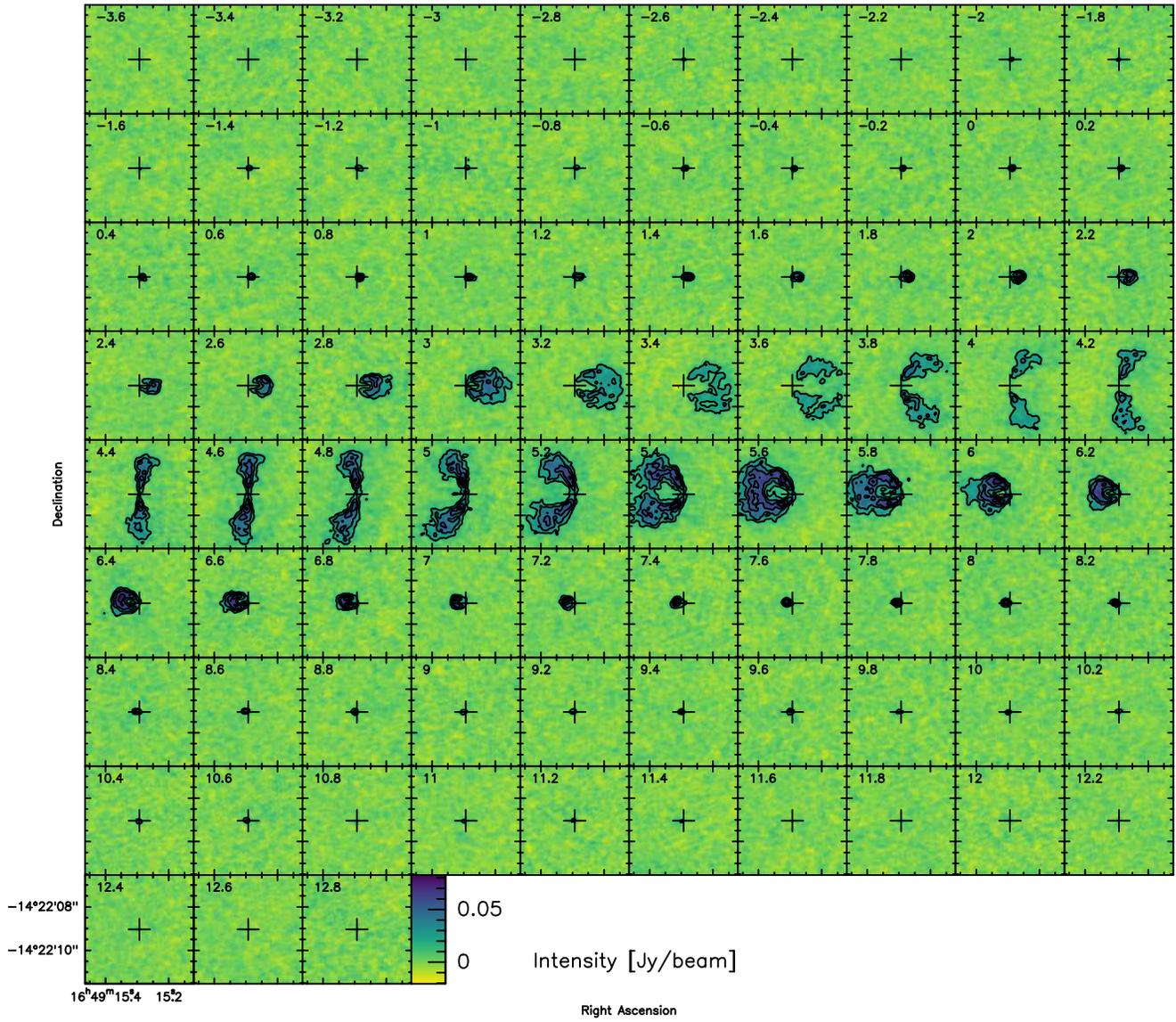}
\caption{CO velocity channel maps. The first contour and the level step are at 5$\sigma$ (where 1$\sigma$=3.5~mJy~beam$^{-1}$). The synthesized beam is 0.25$\arcsec$ $\times$ 0.21$\arcsec$ (PA at -75.9$\degr$). 
 }
\label{fg8}
\end{figure}

\begin{figure}[h!]
\centering
% trim={<left> <lower> <right> <upper>}
\includegraphics[trim={3cm 4.5cm 9cm 1cm},angle=0,width=16cm]{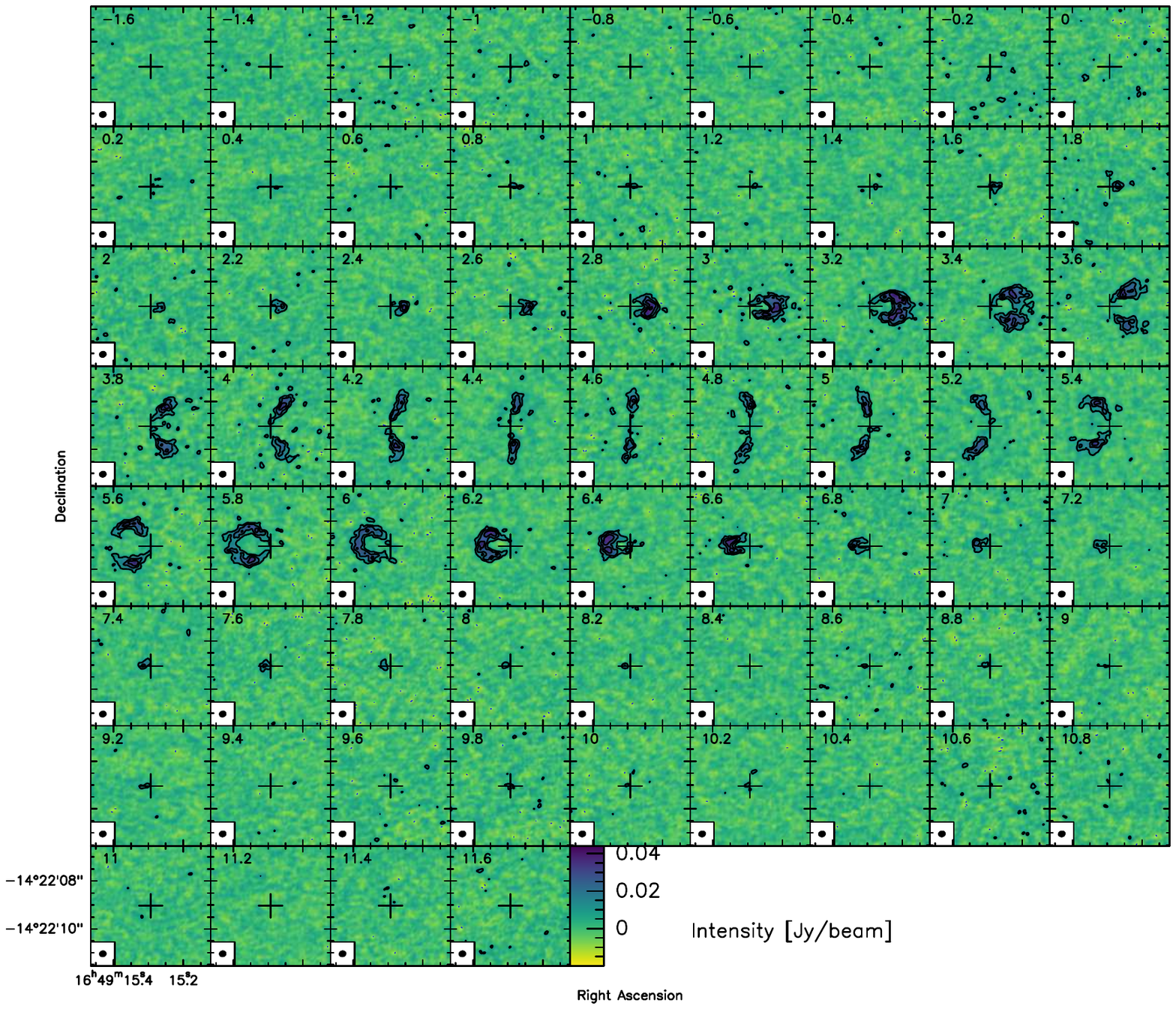}
\caption{$^{13}$CO velocity channel maps. The first contour and the level step are at 3$\sigma$ (where 1$\sigma$=3.4~mJy~beam$^{-1}$). The synthesized beam is 0.25$\arcsec$ $\times$ 0.21$\arcsec$ (PA at -73.4$\degr$).
}
\label{fg9}
\end{figure}

\begin{figure}[h!]
\centering
% trim={<left> <lower> <right> <upper>}
\includegraphics[trim={3cm 4.5cm 9cm 1cm},angle=0,width=16cm]{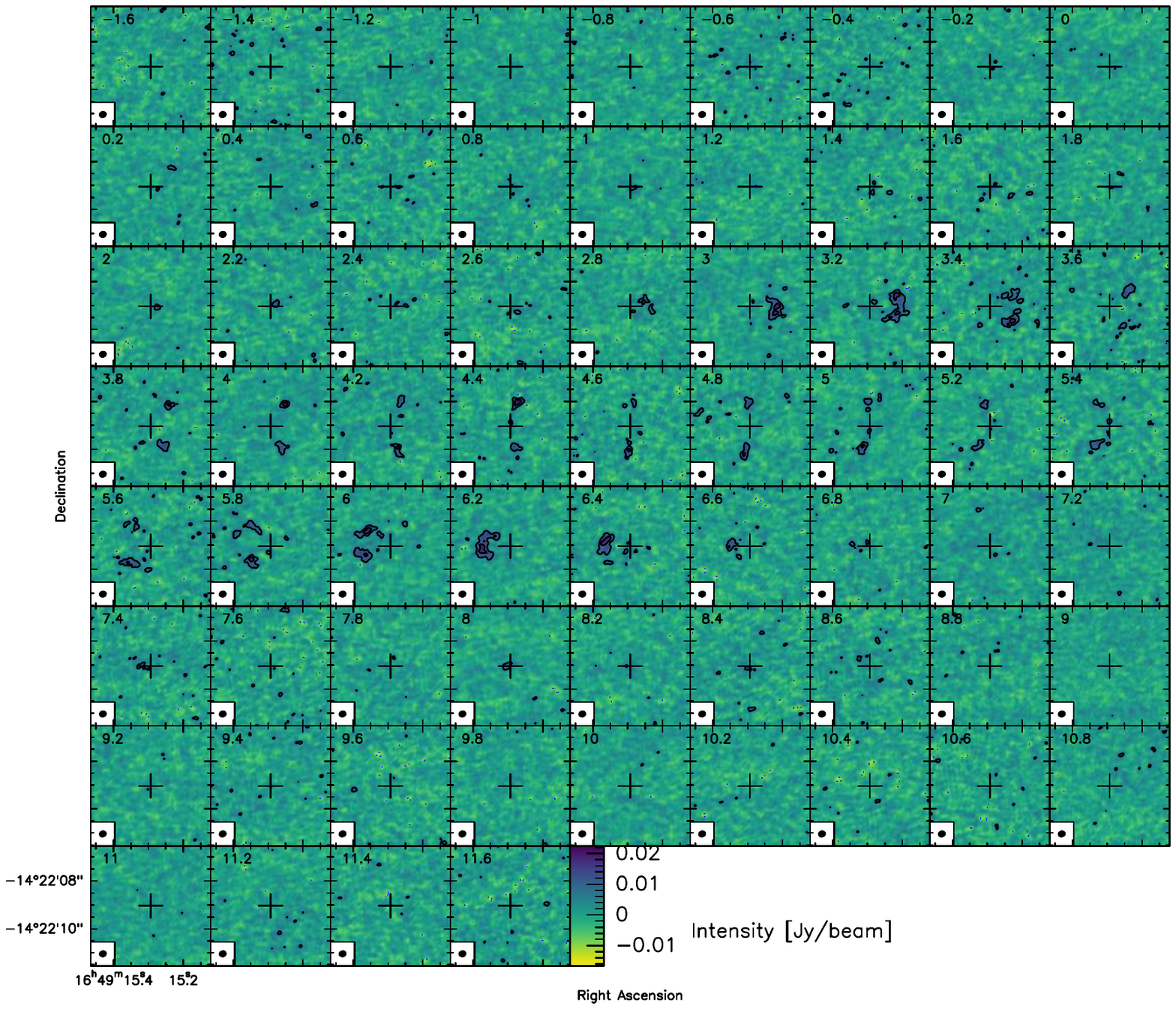}
\caption{C$^{18}$O velocity channel maps. The first contour and the level step are at 3$\sigma$ (where 1$\sigma$=2.6~mJy~beam$^{-1}$). The synthesized beam is 0.26$\arcsec$ $\times$ 0.22$\arcsec$ (PA at -72.4$\degr$).
}
\label{fg10}
\end{figure}

\begin{figure}[h!]
\centering
% trim={<left> <lower> <right> <upper>}
\includegraphics[trim={3cm 2.5cm 9cm 1cm},angle=0,width=16cm]{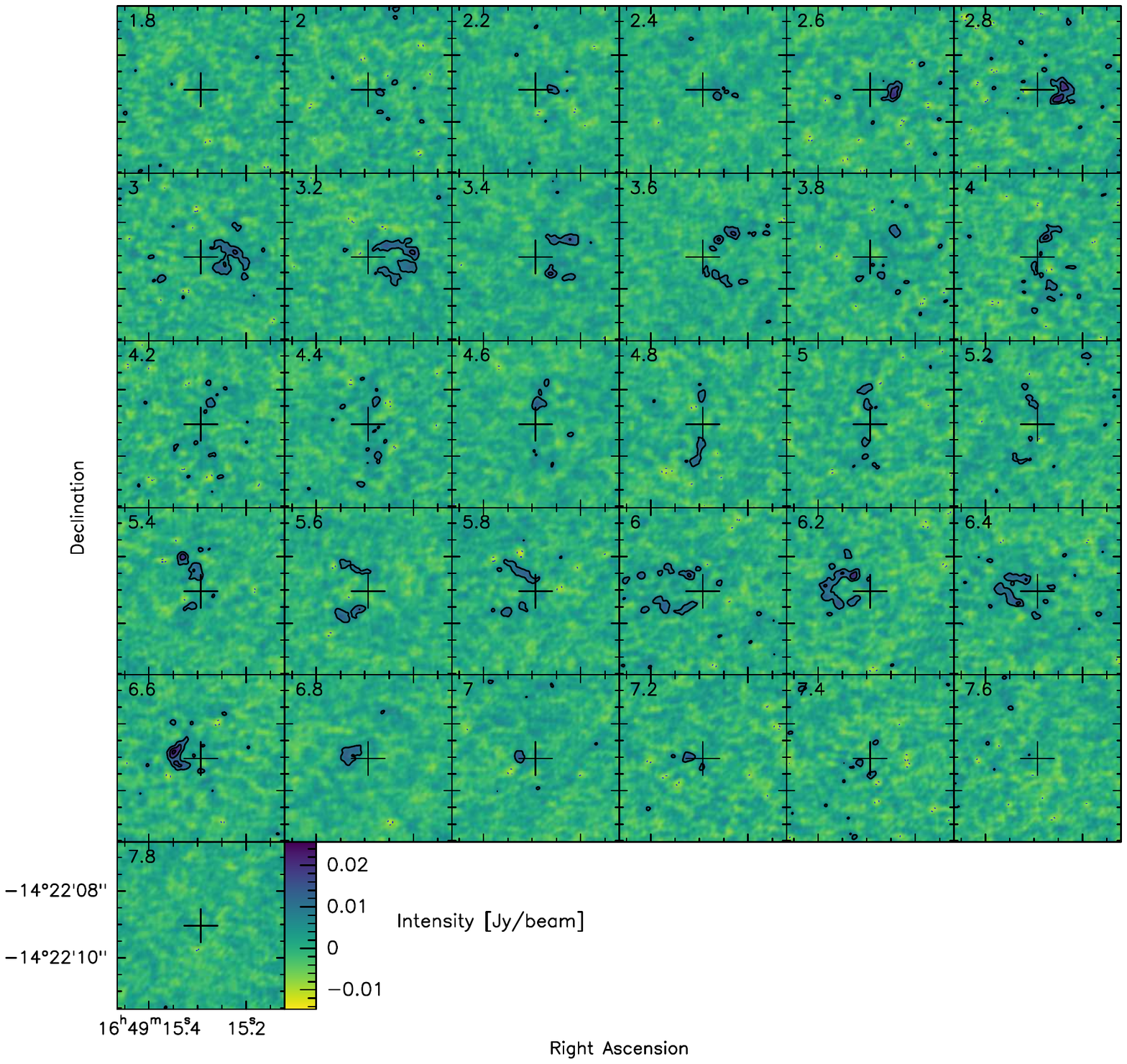}
\caption{DCO$^{+}$ velocity channel maps. The first contour and the level step are at 3$\sigma$ (where 1$\sigma$=2.6~mJy~beam$^{-1}$). The synthesized beam is 0.26$\arcsec$ $\times$ 0.21$\arcsec$ (PA at -73.9$\degr$).}
\label{fg11}
\end{figure}

Figures~\ref{fg8} to \ref{fg11} display the velocity channel maps for $^{12}$CO, $^{13}$CO, C$^{18}$O and DCO$^{+}$, respectively.

\clearpage

%------------------------------------------------------------------
% --- FIGURE 8 ---
%-----------------------------------------------------------------
%

%------------------------------------------------------------------
% --- FIGURE 9---
%-----------------------------------------------------------------
%
%------------------------------------------------------------------
% --- FIGURE 10---
%-----------------------------------------------------------------
%
%------------------------------------------------------------------
% --- FIGURE 11---
%-----------------------------------------------------------------
%

%===================================================================================================
%=============================================
%
%-----------------------------------------------------------------
%------- BIBLIO -------------------
%-----------------------------------------------------------------
%
\clearpage

\bibliographystyle{aasjournal}
%\bibliography{/Users/cecilefavre/Documents/articles/biblio}

%===============================================================
%===============================================================
 \end{document}